\documentclass[aps,prd,nofootinbib,aps,tightenlines,preprintnumbers,notitlepage,longbibliography,superscriptaddress]{revtex4-2}
\usepackage{graphicx}
\usepackage{amsmath}
\usepackage{amsfonts}
\usepackage[colorlinks=true,citecolor=blue,urlcolor=cyan,linkcolor=black]{hyperref}
\usepackage{appendix}
\usepackage{setspace}
\usepackage{simplewick}

\DeclareSymbolFont{toneletters}{T1}{\familydefault}{m}{it}
\SetSymbolFont{toneletters}{bold}{T1}{\familydefault}{bx}{it}

\DeclareMathSymbol\ethm{\mathord}{toneletters}{"F0}

\usepackage{slashed}
\usepackage{bm}

\usepackage{tabularx}
\newcolumntype{C}{>{\centering\arraybackslash}X}
\newcolumntype{R}{>{\raggedleft\arraybackslash}X}

\def\spinup{\partial\kern-0.3em\raise0.42ex\hbox{\tiny\textbackslash}}
\def\spindown{\overline{\partial\kern-0.3em\raise0.42ex\hbox{\tiny\textbackslash}}}

\def\x{{\bf x}}

\def\k{{\bf k}}

\def\l{{\bf l}}

\def\hv{{\hat v}}
\def\th{{\bm\theta}}

\def\nn{\nonumber}
\def\N{{\mathcal N}}

\usepackage[usenames, dvipsnames]{color}
\usepackage[normalem]{ulem}
\usepackage{xcolor}

\newcommand{\dd}{{\rm d}}

\def\bx{{\boldsymbol{x}}}

\newcommand{\btheta}{{\boldsymbol{\theta}}}

\newcommand{\bk}{\boldsymbol{k}}
\newcommand{\bl}{\boldsymbol{l}}
\newcommand{\be}{\begin{eqnarray}}

\newcommand{\ee}{\end{eqnarray}}

\definecolor{colorA}{HTML}{1E90FF}
\definecolor{colorB}{HTML}{228B22}
\definecolor{colorC}{HTML}{FF7F00}
\definecolor{colorD}{HTML}{4B0082}
\definecolor{colorE}{HTML}{B22222}

\definecolor{lgreen}{HTML}{32CD32}
\definecolor{lgray}{HTML}{D3D3D3}
\definecolor{dblue}{HTML}{1E90FF}
\definecolor{orange}{HTML}{FF4500}
\definecolor{indigo}{HTML}{4B0082}

\definecolor{teal}{HTML}{008080}
\definecolor{firebrick}{HTML}{B22222}
\definecolor{salmon}{HTML}{FA8072}
\definecolor{darkgreen}{HTML}{006400}

\usepackage{multirow}
\usepackage{todonotes}

\newcommand{\perimeter}{Perimeter Institute for Theoretical Physics, 31 Caroline St N, Waterloo, ON N2L 2Y5, Canada}

\newcommand{\berkeleya}{Lawrence Berkeley National Laboratory, One Cyclotron Road, Berkeley, CA 94720, USA}
\newcommand{\berkeleyb}{Berkeley Center for Cosmological Physics, Department of Physics, University of California, Berkeley, CA 94720, USA}

\newcommand{\upitta}{Department of Physics and Astronomy, University of Pittsburgh, 3941 O'Hara Street, Pittsburgh, PA 15260, USA}

\newcommand{\upittb}{Pittsburgh Particle Physics, Astrophysics, and Cosmology Center (PITT PACC), University of Pittsburgh, Pittsburgh, PA 15260, USA}

\newcommand{\kavlia}{Kavli Institute for Particle Astrophysics and Cosmology, 382 Via Pueblo Mall Stanford, CA 94305-4060, USA}

\newcommand{\kavlib}{SLAC National Accelerator Laboratory, 2575 Sand Hill Road Menlo Park, California 94025, USA}

\newcommand{\usc}{Physics \& Astronomy Department, University of Southern California, Los Angeles, California, 90089-0484}

\graphicspath{ {plots/} }

\begin{document}

\title{First detection of the moving lens effect with ACT and DESI LS}

\author{Selim~C.~Hotinli}
\email{shotinli@perimeterinstitute.ca} 
\affiliation{\perimeter}

\author{Kendrick~M.~Smith}
\affiliation{\perimeter}

\author{Simone~Ferraro}
\affiliation{\berkeleya}
\affiliation{\berkeleyb}

\author{Ali~Beheshti}
\affiliation{\upitta}
\affiliation{\upittb}

\author{Arthur~Kosowsky}
\affiliation{\upitta}
\affiliation{\upittb}

\author{Elena~Pierpaoli}
\affiliation{\usc}

\author{Emmanuel~Schaan}
\affiliation{\kavlia}
\affiliation{\kavlib}

\begin{abstract}

The moving lens effect is a secondary CMB anisotropy induced by the transverse motion of gravitational potentials.  We develop a Fourier-space cross-spectrum estimator that retains the scale dependence of the signal, and apply it to the Atacama Cosmology Telescope (ACT) DR6 CMB temperature maps and luminous red galaxies from the DESI Legacy Imaging Surveys.  Using the foreground-reduced ACT NILC map, we find strong evidence for a non-zero amplitude of the cross-correlation $b_{\rm ML} = 1.24 \pm 0.26$ ($4.8\sigma$) for the extended sample and $0.93 \pm 0.25$ ($3.7\sigma$) for the main sample, both consistent with the halo-model prediction for the moving lens signal. Our Fourier-based pipeline enforces separation of scales between the reconstructed velocities and the cross-correlation, which we show is essential for foreground mitigation. The residual foreground contamination is expected to be significantly smaller than the signal from both simulations and the multi-frequency analysis presented in this paper. No curl-mode test exceeds $2\sigma$, and the results are robust across analysis variants. They constitute the first detection of the moving lens effect and unlock access to transverse velocities, a new cosmological probe. When combined with the kinematic Sunyaev-Zel'dovich effect, this provides a path toward mapping the three-dimensional velocity field of the Universe, opening a new avenue for probing the growth of structure and gravity on large scales.

\end{abstract}

\maketitle

\section{Introduction}\label{sec:intro}

Peculiar velocities of large-scale structure carry information about the growth rate of cosmic perturbations, the nature of gravity, and the initial conditions of the universe.  The radial component of the velocity field — the projection along the line of sight — has been studied extensively through redshift-space distortions in galaxy surveys~\cite{DESI:2024hhd} and through the kinetic Sunyaev-Zel'dovich (kSZ) effect, which imprints a Doppler shift on CMB photons scattered by free electrons in
motion~\cite{Sunyaev:1980nv, Hand2012, Bernardis2017, ACTPol:2015teu, Smith:2018bpn}.
The \emph{transverse} component — perpendicular to the line of sight — has, by contrast, remained largely inaccessible to observation.  Accessing both components of the velocity field would provide additional avenues for cosmological inference, from breaking the optical-depth degeneracy inherent in kSZ measurements to enable independent constraints on the growth rate $f\sigma_8$ and the amplitude of local primordial
non-Gaussianity $f_{\rm NL}$~\cite{Dalal:2007cu,Hotinli:2021hih}.

The moving lens effect provides a path to measuring transverse velocities from the CMB.  
First discussed by Birkinshaw \& Gull~\cite{1983Natur.302..315B} in 1983 and studied further by Aghanim et al.~\cite{Aghanim:1998ux}, the effect arises because a gravitational potential well moving transverse to the line of sight produces a time-varying potential along the photon path.  
CMB photons traversing such a potential experience a net energy shift, also known as the integrated Sachs-Wolfe effect~\cite{1967ApJ...147...73S,1968Natur.217..511R}, here sourced by bulk transverse motion rather than by the growth or decay of potentials. 
The resulting temperature anisotropy is proportional to the line-of-sight integral of $\boldsymbol{\nabla}_{\!\perp}\Psi \cdot \boldsymbol{v}_\perp$ (Eq.~\ref{eq:main_1}), where $\Psi$ is the Weyl potential and $\boldsymbol{v}_\perp$ is the transverse peculiar velocity.
For a single halo, the signal is a temperature \emph{dipole} oriented along the direction of the halo's transverse motion, with an amplitude proportional to the lensing deflection angle times $v_\perp/c$.  
Two physically equivalent pictures describe the effect: in the CMB rest frame, it is a non-linear ISW contribution from the motion of potentials; in the rest frame of the lens, it is the gravitational lensing of the CMB kinematic dipole as seen by the moving halo~\cite{Hotinli:2018yyc}.

A distinctive feature of the moving lens signal is that it is \emph{purely gravitational}.  
Unlike the kSZ effect, which depends on the electron density, ionization fraction, and optical depth of intervening gas, the moving lens signal depends only on the gravitational potential and the velocity field.  
%complicates the interpretation of kSZ measurements and 
This makes it insensitive to the baryonic physics that introduces an optical-depth degeneracy between the signal amplitude and the electron profile of halos in kSZ measurements~\cite{Smith:2018bpn}.  
The moving lens signal also has a blackbody frequency spectrum, identical to that of the primary CMB, which means it survives internal linear combination (ILC) cleaning but cannot be separated from the primary CMB, kSZ, or halo lensing signals by frequency information alone.

Several studies have developed the theoretical framework for detecting the moving lens effect.
Hotinli et al.~\cite{Hotinli:2018yyc} introduced a quadratic estimator — analogous to standard CMB lensing reconstruction — that cross-correlates small-scale CMB temperature with a density-derived gravitational potential map to reconstruct the large-scale transverse velocity potential, forecasting detection at $\sim$\,$8\sigma$ with the Simons Observatory~\cite{SimonsObservatory:2018koc} combined with DES and $\sim$\,$40\sigma$ with CMB-S4 combined with LSST~\cite{LSSTScience:2009jmu}.  
Hotinli, Johnson \& Meyers~\cite{Hotinli:2020ntd} developed a complementary real-space matched-filter approach, deriving optimal filters for extracting transverse velocities from CMB cutouts around individual halos, and finding consistent signal-to-noise forecasts. 
Yasini, Mirzatuny \& Pierpaoli~\cite{Yasini:2018rrl} proposed a pairwise transverse-velocity estimator — the transverse analog of the pairwise kSZ statistic — as an additional detection pathway.

The cosmological science case for the moving lens was demonstrated by Hotinli et al.~\cite{Hotinli:2021hih}, who showed that transverse velocity reconstruction can measure $f\sigma_8$ to high precision by breaking two degeneracies that limit other probes: the optical-depth degeneracy in kSZ tomography (where $f\sigma_8$ is degenerate with the galaxy-electron cross-power spectrum $P_{ge}$), and the RSD bias in galaxy clustering (where tidal alignment effects introduce a multiplicative bias $b_{\rm rsd}$ on the inferred growth rate~\cite{Obuljen:2020ypy}).  
Like in the case of velocity reconstruction from kSZ, moving-lens tomography reconstructs velocities up to an overall bias $b_\perp$. This bias depends instead on the galaxy-gravitational-potential cross-power spectrum $P_{g\Psi}$, a quantity that can be calibrated with galaxy-lensing measurements, providing an independent and complementary constraint.  
Combined with kSZ and galaxy clustering in a joint analysis, moving lens tomography can improve constraints on $f\sigma_8$ substantially, particularly when astrophysical systematic uncertainties are large.

More recently, significant attention has been given to the practical challenges of extracting the moving lens signal from realistic CMB and galaxy data.
Hotinli et al.~\cite{Hotinli:2023ywh} identified a previously unrecognized potential source of contamination: correlations between the gradients of extragalactic foreground emission (tSZ and CIB) and the transverse velocity field.
Because halos tend to move toward overdense regions, the foreground emission is enhanced ahead of the halo in the direction of its transverse motion, producing a pattern morphologically similar to the moving lens signal, but up to an order of magnitude larger on single-frequency maps.
This foreground signal has a different frequency dependence and spatial profile than the moving lens, offering potential routes to mitigation through multi-frequency component separation or through the scale-dependence of the signal. 

Two independent groups have since performed realistic, map-based forecasts that incorporate correlated non-Gaussian foregrounds.
Hotinli \& Pierpaoli~\cite{Hotinli:2024tjb} used the \textsc{websky} simulation to evaluate both oriented stacking and pairwise estimation methods, finding that residual CIB, tSZ, and halo lensing gradients make the moving lens detection difficult, unless these foregrounds are well understood and/or mitigated. 
Beheshti, Schaan \& Kosowsky~\cite{Beheshti:2024dxw} arrived at broadly consistent conclusions and introduced a template deprojection technique that partially removes the isotropic foreground signal correlated with galaxy positions while preserving the moving lens dipole.  They forecast signal-to-noise ratios ranging from $\sim$\,$2$ for ACT $\times$ DESI~Y1 to $\sim$\,$10$ for SO $\times$ DESI~Y5 and $\sim$\,$100$ for S4 $\times$ LSST with optimal mass weighting, and emphasize that increasing galaxy number counts is more effective than improving CMB sensitivity for detection.
These forecasts suggest that current Stage-3 data may be sensitive to the moving lens signal.

In this work, we present a new approach to measuring the moving lens signal.  
Rather than using the quadratic estimator formalism of \cite{Hotinli:2018yyc} or stacking CMB patches at individual halo locations (the matched-filter approach), we construct a transverse velocity template from a galaxy density field \cite{Guachalla:2023lbx, Hadzhiyska:2023nig}.
Our approach is similar in spirit to the matched-filter analysis of Ref.~\cite{Hotinli:2020ntd}, but retains the scale dependence of the cross-spectrum rather than integrating over $\ell$, which provides direct sensitivity to the scale dependence and frequency dependence of the signal.
We cross-correlate the spin-1 gradient-mode harmonic coefficients with the CMB temperature in harmonic space to measure the angular power spectrum $C_\ell^{TG}$.
This spectrum is proportional to (and provides a direct measurement of) the galaxy-potential cross-power spectrum $P_{g\Psi}(k)$.

The harmonic-space framework provides a clean separation between signal and noise as a function of angular scale and connects directly to the theoretical prediction for $C_\ell^{TG}$.
The normalization is computed using surrogate Gaussian random fields~\cite{Hotinli:2025tul}, a technique that accounts for the survey geometry, selection function, and masks without requiring mock simulations. It also allows a clean separation of scales between the transverse-velocity template and the scales used in the analysis, which we will show is crucial for foreground mitigation.

We apply this estimator to CMB temperature maps from the Atacama Cosmology Telescope (ACT)~\cite{Naess:2020wgi,ACT:2023wcq} cross-correlated with galaxy catalogs from the DESI Legacy Imaging Surveys~\cite{DESI:2016fyo,DESI:2022gle}.
We fit each CMB map independently and use the consistency of $b_{\rm ML}$ across frequencies, together with null tests from frequency-difference maps and curl-mode cross-spectra, to assess foreground contamination.
As a conservative cross-check, we also inspect the cross-correlation of \textit{Planck} 353 and 545 GHz maps~\citep{Planck:2020olo} with the velocity template as a sensitive test of possible CIB contamination, and evaluate the expected foreground biases in a simulation pipeline based on Quijote $N$-body snapshots~\citep{Villaescusa-Navarro:2019bje} and simplified tSZ/CIB models~\cite{Maniyar:2020tzw}.  
Applying this pipeline to ACT DR6 $\times$ DESI LS yields the first observational detection of the moving lens effect at $4.8\sigma$ in the foreground-cleaned NILC map for the extended LRG sample (at $3.7\sigma$ in the main sample).

The remainder of this paper is organized as follows. Section~\ref{sec:prelim} establishes notation, introduces the spin-1 gradient/curl decomposition of the vector fields on the sphere, and defines the moving lens signal. Section~\ref{sec:pipeline} describes the pipeline, including the $P_{g\Psi}$ estimator, surrogate-field normalization, and statistical analysis framework. Section~\ref{sec:data} describes the galaxy catalogs and CMB maps used. Section~\ref{sec:results} presents the $b_{\rm ML}$ measurements from NILC and the single-frequency channels, null tests from frequency-difference and curl-mode cross-spectra, and robustness checks. Section~\ref{sec:foreground} bounds CIB contamination empirically using Planck 353 and 545\,GHz cross-correlations, and evaluates foreground biases in a simulation pipeline based on Quijote $N$-body snapshots. Section~\ref{sec:discussion} discusses the implications and concludes. Technical details of the normalization and the simulation pipeline appear in Appendices~\ref{app:normalization_and_surrogates} and~\ref{app:foreground_sims}.

\section{Preliminaries}\label{sec:prelim}

We denote three-dimensional comoving positions by $\bx$ and two-dimensional sky directions by $\btheta$ (a unit three-vector).  
The comoving distance to redshift $z$ is $\chi(z)$; we will sometimes change variables $\chi \leftrightarrow z$ without comment.
Our Fourier convention is 
\begin{equation}\label{eq:fourier_conv}
f(\x) = \int \frac{d^3\k}{(2\pi)^3} \, \tilde f(\k)\, e^{i\k\cdot\x}\,,
\hspace{1.5cm}
\tilde f(\k) = \int d^3\x \, f(\x)\, e^{-i\k\cdot\x}\,.
\end{equation}
We use the shorthand integral notation (for a fixed 3-d vector ${\bf K}$):
\begin{equation}
\int_{\bk+\bk'={\bf K}} \big( \cdots \big)
  = \int \frac{d^3\bk}{(2\pi)^3} \, \frac{d^3\bk'}{(2\pi)^3} \,
   \big( \cdots \big) \, (2\pi)^3 \delta^3(\bk+\bk'-{\bf K})\,.
\end{equation}
We denote 3-d vector indices by $i,j,\cdots$.
In a 2-d spherical geometry, we denote tangent vector indices by $a,b,\cdots \in \{\theta,\phi\}$.
For example, a 2-d vector field would be denoted $X_a(\btheta)$ in pixel space.
In harmonic space, we use a gradient/curl representation $(X^G_{\ell m}, X^C_{\ell m})$ for vector fields:
\begin{equation}\label{eq:spin1_exp}
(X_\theta \pm iX_\phi)(\btheta)
= \sum_{\ell m}(\mp X^G_{\ell m} - i\,X^C_{\ell m})\,
{}_{\pm 1}Y_{\ell m}(\btheta)\,.
\end{equation}
The gradient coefficients $X^G_{\ell m}$ have parity $(-1)^\ell$ (electric type) and the curl coefficients $X^C_{\ell m}$ have parity $(-1)^{\ell+1}$ (magnetic type).  

The moving lens contribution to the CMB temperature anisotropy is a line-of-sight integral coupling the transverse velocity field to the gradient of the gravitational potential~\cite{1983Natur.302..315B, Aghanim:1998ux}:
\begin{equation}\label{eq:main_1}
\Theta_{\rm ML}(\btheta) = -2 \, T_{\rm CMB} \int\dd\chi\,
\boldsymbol{\nabla}_{\!\perp}\Psi(\bx)\cdot\boldsymbol{v}_\perp(\bx)\,,
\end{equation}
where $\Psi$ is the Weyl potential and $\boldsymbol{v}_\perp$ is the velocity transverse to the line of sight.

\section{Pipeline Description}\label{sec:pipeline}

\subsection{Intuitive idea and simplified pipeline}
\label{ssec:simplified_pipeline}

In this section, we construct a simplified version of our moving-lens estimator, in a flat-sky ``snapshot'' geometry with no survey mask or lightcone evolution, and build intuition by emphasizing similarity with kSZ estimators.
In the next section (\S\ref{ssec:curved_sky}), we relax these simplifying assumptions and construct an estimator that can be applied to real data.

In the snapshot geometry, large-scale structure fields are defined in a 3-d periodic box of side length $L$, at a fixed redshift $z_*$.
The CMB is a 2-d flat-sky field obtained by projecting onto one periodic face of the box, and switching to angular coordinates $\btheta = (\bx^\perp/\chi_*)$.

{\bf Review of kSZ Fourier-space estimator.}
Our moving lens estimator is very similar to estimators that measure the galaxy-electron power spectrum $P_{ge}(k)$ using the kSZ effect, especially the Fourier-space estimators from \cite{Harscouet:2025pwl,Qu:2026zyh, Hadzhiyska:2026uyq}.
Roughly speaking, the moving lens estimator is the ``kSZ estimator with more indices'': some scalar quantities become vector fields.
We briefly review kSZ estimators, before constructing a moving-lens estimator.

Let $\delta_g(\x)$ be a galaxy field with 3-d comoving number density $n_g^{3d}$, and let $\hv_j(\x)$ be a large-scale velocity reconstruction derived from $\delta_g$.
In this paper, we use a linear reconstruction:
\begin{equation}
\hv_j(\k) = (ik_j) U(k) \delta_g(\k) 
  \hspace{1cm} \mbox{where }
      U(k) \equiv
    \begin{cases}
    \displaystyle\frac{faH}{k^2b_g}
      \frac{P_{gg}(k)}{P^{\mathrm{tot}}_{gg}(k)}  & k < k_{\mathrm{max}}\,,
      \\[6pt]
    0 & \text{otherwise}\,.
    \end{cases}
    \label{eq:vhat_flat}
\end{equation}
where $P_{gg}^{\rm tot}(k), P_{gg}(k)$ denote the galaxy power spectrum with and without Poisson noise.
However, the precise form of the velocity reconstruction will not be very important.
To quantify the fidelity of the reconstruction, we define parameters $\eta_r, \eta_\perp$ by:
\begin{equation}
\big\langle v_r^{\rm true}(\x) \, \hv_r(\x) \big\rangle = \eta_r\,,
  \hspace{1.5cm}
\big\langle v_{a}^{\rm true}(\x) \, \hat v_{b}(\x) \big\rangle = \frac{\eta_\perp}{2} \, \delta_{ab}\,,
\label{eq:eta_def_flat}
\end{equation}
where $v_r$ denotes the radial component of the velocity field\footnote{In the kSZ literature, a quantity closely related to $\eta_r$ is typically used to normalize the estimators, that is, the correlation coefficient between true and reconstructed radial velocities $r_v \equiv \eta_r / [\sigma(v_r^{\rm true}) \sigma(\hat v_{r})]$.}, and $a,b\in\{1,2\}$ denotes transverse components. 

In the kSZ case, we construct a 2-d field $\pi(\btheta)$ by summing galaxies weighted by radial velocities:
\begin{equation}
\pi(\btheta) = \sum_{i\in \rm gal} \hv_r(\x_i) \, \delta^2(\btheta-\btheta_i) \label{eq:pi_def}\,.
\end{equation}
Then, it can be shown that the kSZ effect produces a correlation $C_l^{T\pi} \ne 0$ between the CMB temperature and the $\pi$-field:
\begin{equation}
C_l^{T\pi} = \eta_r n_g^{3d} K_* L \, P_{ge}(l/\chi_*)
  \hspace{1.5cm} \mbox{where }
  K_* \equiv -T_{\rm CMB} \sigma_T n_{e0} \, (1+z_*)^2
  \label{eq:cltpi}\,.
\end{equation}
We omit the derivation of Eq.\ (\ref{eq:cltpi}), since we do the analogous calculation for the moving-lens effect in detail below.
Thus, the estimator $C_l^{T\pi}$ can be used to detect the kSZ effect, and measure $P_{ge}(k)$ as a function of $k$.

{\bf Moving-lens estimator.}
How should the preceding construction be modified for the moving lens effect?
Intuitively, instead of weighting galaxies by radial velocities (as in Eq.\ (\ref{eq:pi_def})), we want to weight them by transverse velocities.
Thus, we get a vector (spin-one) field $X_a(\btheta)$, instead of a scalar (spin-zero) field $\pi(\btheta)$:
\begin{equation}
X_a(\btheta) = \sum_{i\in \rm gal} \hv_a(\x_i) \, \delta^2(\btheta-\btheta_i) \label{eq:Xa_flat}\,.
\end{equation}
We decompose $X_a$ into its gradient and curl modes $X_G,X_C$.
This step is the flat-sky analog of Eq.\ (\ref{eq:spin1_exp}), and is naturally written in Fourier space:
\begin{equation}
X_a(\bl) = \frac{il_a}{l} G(\bl) + \frac{i\epsilon_{ab} l_b}{l} C(\bl)\,.
\end{equation}
We claim that the moving lens effect produces a correlation $C_l^{TG} \ne 0$ between the CMB temperature and the {\bf gradient mode of the $X$-field.}
The rest of this subsection is devoted to deriving this result.

First, we write $T_{ML}(\btheta)$ and $X_a(\btheta)$ (Eqs.\ (\ref{eq:main_1}), (\ref{eq:Xa_flat})) in flat-sky Fourier space:
\begin{align}
T_{ML}(\bl) &= - 2\frac{T_{\rm CMB}}{\chi_*^2}
  \int_{\bk_L+\bk_S=\bl/\chi_*}
  v_b^{\rm true}(\k_L) \, \big( \partial_b\Psi(\k_S) \big)\,, 
  \label{eq:Tflat_fourier} \\
X_a(\bl) &= n_g^{3d} \int_{\bk_L+\bk_S=\bl/\chi_*}
  \hv_a(\bk_L) \, \delta_g(\bk_S)\,.
  \label{eq:Xflat_fourier}
\end{align}
Using these expressions, we compute the two-point function $\langle T X_a \rangle$, using a sequence of ``squeezed'' approximations which are familiar from the kSZ context:
\begin{align}
\big\langle T_{ML}(\bl) X_a(\bl') \big\rangle
  &\sim 
  \contraction[2.5ex]{\Big( v_b^{\rm true} \big( \partial_b}{\Psi}{\big) \Big) \Big( \hv_a}{\delta_g}
  \contraction{\Big(}{v_b^{\rm true}}{\big( \partial_b \Psi \big) \Big) \Big(}{\hv_a}
  \Big( v_b^{\rm true} \big( \partial_b \Psi \big) \Big) \Big( \hv_a \delta_g \Big)
  \hspace{5cm} \mbox{(schematic)} \nn \\
&= - 2\frac{n_g^{3d} T_{\rm CMB}}{\chi_*^2}
 \int_{\substack{\bk_L+\bk_S=\bl/\chi_* \\ \bk'_L+\bk'_S=\bl'/\chi_*}}
\big\langle v_b^{\rm true}(\bk_L)
 \hv_a(\bk'_L) \big\rangle \,
 (i\k_{Sb}) 
\big\langle \Psi(\k_S) \delta_g(\k_S') \big\rangle \nn \\
&\approx - 2\frac{n_g^{3d} T_{\rm CMB}}{\chi_*^2}
\left( \frac{\eta_\perp}{2} \delta_{ab} \right)
\left( \frac{i\bl_b}{\chi_*} \right) 
P_{g\Psi}\bigg( \frac{l}{\chi_*} \bigg) \,\,
L \chi_*^2 \, (2\pi)^2 \delta^2(\bl+\bl') \nn \\
&= {\mathcal N} \, (i\bl'_a) \,
P_{g\Psi}\bigg( \frac{l}{\chi_*} \bigg) \,
(2\pi)^2 \delta^2(\bl+\bl') 
\hspace{3cm}
  \mbox{where } {\mathcal N} \equiv
   \frac{\eta_\perp \, n_g^{3d} T_{\rm CMB} L}{\chi_*}\,.
  \label{eq:toy_N}
\end{align}
In the first line, we made the approximation that the velocity reconstruction only has fluctuations on scales $k_L \ll (l/\chi_*)$, so that the four-point function on the RHS is well-approximated by the indicated Wick contraction.
In the second and third lines, we evaluated the Wick contraction explicitly in the approximation $\bk_S \approx (\bl/\chi_*)$.
From this calculation, we can read off the power spectra in the gradient/curl basis:
\begin{equation}
C_\ell^{TG} = {\mathcal N} \ell \, P_{g\Psi}(\ell/\chi_*)\,,
  \hspace{1.5cm}
C_\ell^{TC} = 0\,.
 \label{eq:cltg_cltc_flat}
\end{equation}
Note that $C_\ell^{TC}=0$ can also be seen directly from symmetry, since the moving-lens effect is parity-symmetric.

This concludes our derivation that the moving lens effect produces a correlation $C_\ell^{TG} \ne 0$ with the gradient mode of the $X$-field, as claimed above.
The correlation is proportional to $P_{g\Psi}(\ell/\chi_*)$, and the normalization ${\mathcal N}$ is given by Eq.\ (\ref{eq:toy_N}).
Analogously to the kSZ case, the estimator $C_\ell^{TG}$ can be used to detect the moving-lens effect and to measure $P_{g\Psi}(k)$ as a function of $k$.

The galaxy--potential cross-power spectrum $P_{g\Psi}(k)$ is related to the galaxy--matter power spectrum $P_{gm}(k)$ via the Poisson equation:
\begin{equation}\label{eq:PgPsi}
    P_{g\Psi}(k,z) = -\frac{3\,H_0^2\,\Omega_{m0}}{2\,a\,k^2}\,P_{gm}(k,z)\,.
\end{equation}
Since $P_{gm}(k) > 0$, both $P_{g\Psi}(k)$ and $C_\ell^{TG}$ are negative.

We compute a fiducial galaxy--matter power spectrum $P_{gm}^{\rm fid}(k)$ using the \textsc{hmvec} halo-model code~\cite{Smith:2018bpn,Madhavacheril:2019buy}, at the effective redshift $z_*=0.734$. We use the default \textsc{hmvec} HOD, abundance-matched to the 3-d number density of the DESI-LS galaxies. This results in a slightly different $P_{gm}^{\rm fid}(k)$ for the main and extended DESILS LRG samples (see \S\ref{sec:data_gal}).
The resulting predictions for DESI-LS quantities such as the mean halo bias agree at the $\sim 10\%$ level. The shape of $P_{g\Psi}(k)$ is well-constrained by this procedure; the absolute amplitude carries somewhat larger modeling uncertainty since it's proportional to the mean halo mass, rather than the bias. In future work, it can be calibrated externally with galaxy or CMB lensing measurements~\cite{Hadzhiyska:2025mvt}.

\subsection{Curved-sky pipeline and surrogate normalization}
\label{ssec:curved_sky}

In this subsection, we describe our pipeline for detecting the moving lens effect in ACT and DESILS.
The intuitive idea is the same as our simplified pipeline in \S\ref{ssec:simplified_pipeline}, but we now incorporate the complications of real data: an angular mask, redshift evolution, curved-sky geometry.
Many of the details follow our previous paper \cite{Hotinli:2025tul}, which constructed kSZ (radial velocity) estimators for the same datasets (ACT and DESILS-LRG).

We surround the 3-d DESILS footprint with a periodic bounding box, padded to mitigate artifacts from periodic boundary conditions. Three-dimensional fields are represented in Cartesian coordinates in the bounding box, or in harmonic space via 3-d FFTs. Two-dimensional fields are represented as curved-sky {\tt pixell} maps, or in harmonic space via spherical transforms.

First, we define the weighted galaxy overdensity $\rho_g(\x)$ by the standard
galaxies-minus-randoms prescription:
\begin{equation}\label{eq:rho_g}
    \rho_g(\bx) = \sum_{i\in\mathrm{gal}} W_i\,\delta^3(\bx - \bx_i)
    - \frac{N_g}{N_r}\sum_{j\in\mathrm{rand}} W_j\,\delta^3(\bx - \bx_j)\,,
\end{equation}
where $W_i$ is a per-object weight, and $N_g = \sum_i W_i$, $N_r = \sum_j W_j$ are the weighted galaxy and random counts.
This definition of $\rho_g(\x)$ accounts for the survey geometry (via the random catalog), and includes a per-object weight $W_i$ to encode completeness, FKP weighting, and other complications.
In our DESILS pipeline, we choose $W_i$ to downweight galaxies with large photo-$z$ errors (see Eq.\ (\ref{eq:weight_def}) below).

Second, we compute the velocity reconstruction $\hv_j(\x)$ via a linear filtering operation similar to Eq.\ (\ref{eq:vhat_flat}), but using the galaxy density $\rho_g$ defined above:
\begin{equation}
\label{eq:U_filter}
\hv_j(\bx) = \int \frac{d^3\k}{(2\pi)^3} \, (ik_j) U(k) e^{i\k\cdot\x} \, \rho_g(\bk)
\hspace{1cm} \mbox{where }
    U(k) \equiv
    \begin{cases}
    \displaystyle\frac{faH}{k^2b_g}
      \frac{P_{gg}(k)}{P^{\mathrm{tot}}_{gg}(k)}\,, & k < k_{\mathrm{max}}\,,
      \\[6pt]
    0\,, & \text{otherwise}\,.
    \end{cases}
\end{equation}
Here, $f$ is the linear growth rate, $a$ the scale factor, $H$ the Hubble
parameter, $b_g$ the linear galaxy bias, and
$P_{gg}^{\mathrm{tot}} = P_{gg} + 1/\bar{n}_g^{\rm 3d}$ includes shot noise.  The
prefactor $ik_j faH/(k^2b_g)$ converts the galaxy density contrast to a velocity
via the continuity equation, while $P_{gg}/P_{gg}^{\mathrm{tot}}$ is a
Wiener filter that suppresses noise-dominated modes.
All redshift-dependent quantities in Eq.\ (\ref{eq:U_filter}), such as $f, b_g, \bar n_g^{3d}$, are evaluated at the central redshift $z_*=0.734$ of DESILS.

Third, we compute a moving-lens template field $X_a(\btheta)$ by:
\begin{equation}\label{eq:Xa_def}
    X_a(\btheta) = \sum_{i\in\mathrm{gal}} W_i\,
    \hat{v}_a(\bx_i)\,\delta^2(\btheta - \btheta_i)\,-\frac{N_g}{N_r} \sum_{j\in\mathrm{rand}} W_j\,
    \hat{v}_a(\bx_j)\,\delta^2(\btheta - \btheta_j)\,,
\end{equation}
where the notation $\hat v_a$ (as opposed to $\hat v_j$) means that we project onto the transverse (to $\x$) directions $a \in \{\theta, \phi\}$ (see \S\ref{sec:prelim}).
The definition of $X_a(\btheta)$ is similar to Eq.\ (\ref{eq:Xa_flat}) in our simplified pipeline, but we include a per-galaxy weight $W_i$ (see above), and $X_a(\btheta)$ is now a curved-sky 2-d field.

When we decompose $X_a(\btheta)$ into gradient/curl components $X^G_{\ell m}$, $X^C_{\ell m}$, we use the curved-sky decomposition in Eq.~(\ref{eq:spin1_exp}). We then compute:
\begin{equation}
C_\ell^{TG} = \frac{1}{2\ell+1} \sum_m X^G_{\ell m} T_{\ell m}^*\,,
\end{equation}
with no $f_{\rm sky}$ correction on the RHS (that is, $C_\ell^{TG}$ always denotes a ``pseudo'' power spectrum in this paper).
As in our simplified pipeline, the statistic $C_\ell^{TG}$ can be used to detect the moving lens effect, or to measure the power spectrum $P_{g\Psi}(k)$.

However, in our full (non-simplified) pipeline, the relation between $C_\ell^{TG}$ and $P_{g\Psi}(k)$ turns out to be more complicated. In Appendix \ref{app:normalization_and_surrogates}, we'll show that:
\begin{equation}\label{eq:ClTG}
C_\ell^{TG} = \mathcal{N}\,\ell\,b_\ell\,P_{g\Psi}(\ell/\chi_*)
  \hspace{1.5cm} \mbox{where }
    \mathcal{N} \equiv \frac{T_{\mathrm{CMB}}}{4\pi}
    \sum_{j\in\mathrm{gal}} W_j \,
    \frac{\eta_\perp(\bx_j)}{\chi_j^3}\,.
\end{equation}
where a CMB beam $b_\ell$ has been included, and the 3-d field $\eta_\perp(\x)$ is defined by:
\begin{equation}
\label{eq:eta_def}
\eta_\perp(\bx) \equiv \left\langle \sum_{a\in\{\theta,\phi\}} \hat{v}_a(\bx)\,v_a^{\rm true}(\bx) \right\rangle\,.
\end{equation}
Note that $\eta_\perp(\x)$ is a 3-d field that depends on both redshift (due to evolution) and angular location (due to boundary effects from the survey mask).
This definition of $\eta_\perp(\x)$ generalizes the scalar quantity $\eta_\perp$ in the simplified pipeline (Eq.\ (\ref{eq:eta_def_flat})).

The relation (\ref{eq:ClTG}) between $C_\ell^{TG}$ and $P_{g\Psi}(k)$ is nontrivial to derive, and we defer the details to Appendix \ref{app:normalization_and_surrogates}. In this appendix, we also show how to compute $\eta_\perp(\x)$ and ${\mathcal N}$ using a Monte Carlo procedure (``surrogate fields'').
The technical challenge of computing ${\mathcal N}$ is somewhat tangential for purposes of this paper, since ${\mathcal N}$ is not needed for assessing detection significance, null test failure, or the ``shape'' in $\ell$ of the fiducial moving-lens signal.
The value of ${\mathcal N}$ is only needed to compare the amplitude of $P_{g\Psi}(k)$ to a fiducial model. (Note that the amplitude has large modeling uncertainty -- see the discussion at the end of \S\ref{ssec:simplified_pipeline}.)

\subsection{Statistical analysis}\label{sec:statmethod}

After estimating $C_\ell^{TG}$, we bin the power spectrum in $\ell$, obtaining a ``data vector'' $d_b$, where $1 \le b \le N_b$ indexes an $\ell$-bin.
We use the following binning procedure:
\begin{equation}
d_b = \frac{1}{W_b} \sum_{\ell\in b} w_\ell D_\ell\,,
  \label{eq:l_binning}
\end{equation}
where $D_\ell \equiv \ell^2 C_\ell^{TG}$, and the $\ell$-weighting within a bin is $w_\ell \equiv (2\ell+1)$, and $W_b \equiv \sum_{\ell\in b} w_\ell$.
We use $N_b=6$ logarithmically-spaced bins spanning $\ell\in [2500,6000]$.

We estimate the covariance $C_{bb'} = \mbox{Cov}(d_b,d_{b'})$ from the scatter in $D_\ell$ between different values of $\ell$.
A slight complication is that on the partial sky, mode coupling correlates nearby multipoles over a range $\Delta\ell \sim 1/f_{\rm sky}$.
This motivates the following ``empirical'' estimate of the covariance matrix:
\begin{equation}\label{eq:cov_subbinned}
C_{bb'} =
    \frac{1}{W_b\,W_{b'}}\,
    \sum_{\substack{\ell \in b,\, \ell' \in b' \\
    |\ell - \ell'| \le \Delta\ell}}
    w_\ell\,w_{\ell'}\,
    D_\ell \,D_{\ell'} \,.
\end{equation}
This is an unbiased estimator of $\mbox{Cov}(d_b,d_{b'})$, under the assumptions that $\mbox{Cov}(D_\ell,D_{\ell'})=0$ for $|\ell-\ell'| > (\Delta \ell)$, and that $\langle D_\ell \rangle = 0$ (as appropriate for the null hypothesis of no moving-lens detection).
Note that off-diagonal correlations are either zero (if $|b-b'|>1$) or very small (if $|b-b'|=1$).
For the DESI-LS LRG footprints, $f_{\rm sky} \approx 0.09$ (NGC) and $0.19$ (SGC),
giving $\Delta\ell = 11$ and $5$, respectively.

As a check on the empirical estimator (\ref{eq:cov_subbinned}) for the power spectrum covariance, we also implemented a different method based on simulations.
We find that agreement between the two methods is excellent.
The simulation-based method is described in Appendix \ref{app:sim_errors}.

Throughout the paper, we quantify the moving-lens detection significance by fitting the data vector $d_b$ to an overall multiple (denoted $b_{\rm ML}$) of a moving lens template $t_b$.
To obtain $t_b$, we start with the fiducial $P_{g\Psi}^{\rm fid}(k)$ described at the end of \S\ref{ssec:simplified_pipeline}, convert to $C_\ell^{TG} = \mathcal{N}\,\ell\,b_\ell\,P_{g\Psi}(\ell/\chi_*)$, and bin in $\ell$ using Eq.\ (\ref{eq:l_binning}).
The likelihood function for $b_{\rm ML}$ is given by:
\begin{equation}\label{eq:loglike}
    \ln\mathcal{L}(b_{\rm ML}) =
    -\frac{1}{2}\,
    (\boldsymbol{d} - b_{\rm ML}\,\boldsymbol{t})^T\,
    \mathbf{C}^{-1}\,
    (\boldsymbol{d} - b_{\rm ML}\,\boldsymbol{t})\,.
\end{equation}
Maximizing the Gaussian likelihood, we find that the best-fit $b_{\rm ML}$ value and its variance are given by: 
\begin{equation}\label{eq:bml_fit} 
\hat b_{\rm ML} = \frac{\boldsymbol{t}^T\,\mathbf{C}^{-1}\,\boldsymbol{d}} {\boldsymbol{t}^T\,\mathbf{C}^{-1}\,\boldsymbol{t}}\,, \hspace{1.5cm} \sigma^2(b_{\rm ML}) = \frac{1}{\boldsymbol{t}^T\,\mathbf{C}^{-1}\,\boldsymbol{t}}\,. 
\end{equation} 
We use Eq.\ (\ref{eq:bml_fit}) throughout and quote $\hat b_{\rm ML}$ and $\sigma$ as the best-fit value and uncertainty, respectively. When we analyze the NGC+SGC jointly, we add the log-likelihoods (assuming statistical independence). Note that the NGC and SGC have the same fiducial $P_{g\Psi}(k)$, but different fiducial $C_\ell^{TG} = \mathcal{N}\,\ell\,b_\ell\,P_{g\Psi}(\ell/\chi_*)$, since the normalization $\N$ is different in the NGC and SGC (roughly as $\N \propto f_{\rm sky}$).

To quote detection significance, we define the matched-filter signal-to-noise ratio:
\begin{equation}\label{eq:snr_def}
    \mathrm{SNR} = \frac{\hat b_{\rm ML}}{\sigma(b_{\rm ML})} =
    \frac{\boldsymbol{t}^T\,\mathbf{C}^{-1}\,\boldsymbol{d}}
    {\sqrt{\boldsymbol{t}^T\,\mathbf{C}^{-1}\,\boldsymbol{t}}}\,,
\end{equation}

\subsection{Biases from CMB lensing, kSZ, and foregrounds}

One might worry that ordinary (non-moving) CMB lensing would be a large contaminant to the moving-lens signal. However, there is a symmetry argument which shows that the lensing bias to $C_\ell^{TG}$ vanishes. Consider a symmetry which reverses the sign of the primary CMB temperature $T_{\rm pri} \rightarrow (-T_{\rm pri})$ on the last scattering surface, leaving large-scale structure in the late universe unmodified. The CMB lensing contribution to our moving-lens estimator $C_\ell^{TG}$ is odd under this symmetry. Therefore, the mean bias $\langle C_\ell^{TG} \rangle$ due to CMB lensing must be zero.

Going beyond mean bias, one may wonder whether statistical errors on $C_\ell^{TG}$ are affected by the non-Gaussian statistics of CMB lensing.
Here, we note that the estimator $C_\ell^{TG}$ involves one power of CMB temperature, so its covariance $\mbox{Cov}(C_\ell^{TG}, C_{\ell'}^{TG})$ depends only on the two-point function $C_\ell^{TT}$, not on higher-point statistics.
Therefore, the effect of CMB lensing is no different from other contributions to the temperature power spectrum, and is fully captured by our method (\S\ref{sec:statmethod}) for assigning error bars.

Similarly, one might worry that kSZ is a potential contaminant.
Here, a different symmetry applies: radial reflection $v_r \rightarrow (-v_r)$ in the line-of-sight direction.
(Note that this is an exact symmetry in the Limber/snapshot approximation, but an approximate symmetry in an evolving lightcone geometry.)
The kSZ contribution to $C_\ell^{TG}$ is odd under this symmetry, so the same arguments as in the CMB lensing case apply, to show that kSZ contamination is not an issue.

Astrophysical foregrounds (tSZ, CIB, radio sources) are a different story: here there is no symmetry in sight, and foreground bias to the moving-lens estimator is a concern.
Indeed, several previous studies~\cite{Hotinli:2023ywh, Hotinli:2024tjb, Beheshti:2024dxw} have found significant foreground biases to the moving-lens signal, and highlighted their importance.
For this reason, our results and null tests in subsequent sections (\S\ref{sec:results}, \S\ref{sec:foreground}) are largely aimed at quantifying foreground biases.

One of the main results of this paper is that CMB foreground biases to the moving-lens estimator are small, provided that the estimator is constructed in a particular way (Fourier-space throughout, hard $k_{\rm max}$ cutoff in the velocity reconstruction).
We'll show this in two ways: either via frequency-difference null tests applied to data (\S\ref{sec:nulltests},\S\ref{ssec:planck_353_545}), or via a simplified moving-lens pipeline applied to Quijote simulations (\S\ref{ssec:quijote_foregrounds}).

\section{Data}\label{sec:data}

We use the same galaxy catalog and CMB maps as in the kSZ
analysis in Ref.~\cite{Hotinli:2025tul}.  This section summarizes the key
properties of each data set and the preprocessing steps specific to
the moving lens analysis.

\subsection{Galaxy catalog}\label{sec:data_gal}

For the galaxy survey, we use the luminous red galaxy (LRG) sample
from the DESI Legacy Imaging Surveys Data Release~9
(DESI-LS~DR9)~\cite{Dey:2018,DESI:2022gle,Zhou:2023gji}.  The DESI
Legacy Imaging Surveys combine data from three optical programs ---
DECaLS, BASS, and MzLS --- covering approximately
$14\,000\;\mathrm{deg}^2$ of extragalactic sky~\cite{Dey:2018}.  The `main' LRG target selection of
Ref.~\cite{Zhou:2023gji} applies the strictest color and
magnitude cuts, providing the most uniform photometry and the highest
photometric redshift purity.

We apply the veto masks and quality cuts described in
Ref.~\cite{White:2021yvw} (see also Section~3.3 of
Ref.~\cite{Zhou:2023gji}).  We analyze both the northern Galactic cap
(NGC) and southern Galactic cap (SGC), as well as their combination.  We further restrict to
the photometric redshift range $0.4 \le z_{\rm obs} \le 1.0$.

After all cuts and after restricting to sky regions where the CMB
pixel weight $W_{\rm CMB}(\btheta) > 0$
(see Section~\ref{sec:data_cmb} below), the main LRG catalog
contains $1{,}672{,}848$ galaxies in NGC
($f_{\rm sky} \approx 0.09$, effective area
$\approx 3{,}670\;\mathrm{deg}^2$),
$3{,}410{,}551$ in SGC ($f_{\rm sky} \approx 0.19$,
$\approx 7{,}710\;\mathrm{deg}^2$).
The NGC+SGC combination is performed at the bandpower level
using inverse-variance weighting of independent per-cap fits
(Section~\ref{sec:statmethod}).
The corresponding random catalogs we use contain $29{,}610{,}244$ (NGC)
and $61{,}888{,}997$ (SGC) objects.

\paragraph{Extended sample.}
Throughout our figures, we display results from the extended sample with the `extended'
DESI-LS~DR9 LRG target selection~\cite{Zhou:2023gji}, which relaxes
the color and magnitude cuts, providing a higher number density at
the cost of slightly larger photometric redshift errors.
The extended catalog contains $4{,}349{,}228$ (NGC),
$8{,}839{,}449$ (SGC) galaxies
over the same footprint as the main sample (the random catalogs are
identical).  Main and extended share the same sky coverage, with the
extended sample adding fainter galaxies; the two are therefore not
independent, and the two results serve as a consistency check
rather than independent datasets.

\paragraph{Photometric redshifts.}
The main LRG sample has photometric redshift errors at the
$2$--$3\%$ level (normalized median absolute deviation
$\mathrm{NMAD} \approx 0.02$--$0.03$), with an outlier fraction
below $0.5\%$~\cite{DESI:2022gle,Zhou:2023gji}.

The per-galaxy weights $W_i$ that appear in the density field
(Eq.~\ref{eq:rho_g}) and in the velocity template
(Eq.~\ref{eq:Xa_def}) are set to
\begin{equation}\label{eq:weight_def}
    W_i = \exp\!\Bigl[-\frac{\sigma_{z,i}^2}
    {\alpha\,(1+z_i)^2}\Bigr]\,,
    \qquad \alpha = 2.5\times 10^{-3}\,,
\end{equation}
where $\sigma_{z,i}$ is the photometric redshift error of galaxy~$i$.
This choice down-weights objects whose photo-$z$ errors are large
compared to the correlation length of the velocity field.

\paragraph{Random catalog.}
For the random catalog, we assign photometric redshifts by
deconvolving the joint $(z_{\rm obs}, \sigma_z)$ distribution of
the galaxy sample, following the procedure described in
Ref.~\cite{Hotinli:2025tul}.  Each random object is assigned a
$(z_{\rm obs}, z_{\rm true}, \sigma_z)$ drawn from the
inferred three-dimensional distribution, and the same redshift and quality cuts are applied.  The random catalog is used both in the density field construction (Eq.~\ref{eq:rho_g}) and in the surrogate normalization (Appendix~\ref{app:normalization_and_surrogates}).

These are the same samples used in kSZ measurements of stacked gas profiles \cite{Hadzhiyska:2024qsl}, as well as used for radial velocity reconstruction \cite{Hotinli:2025tul}, thus allowing joint kSZ and ML analyses in the future.

\subsection{CMB maps}\label{sec:data_cmb}
We use the component-separated NILC temperature map from ACT
Data Release 6 (DR6)~\cite{ACT:2023wcq}, which combines ACT
(DR4 and DR6, 93--225\,GHz) and Planck
NPIPE~\cite{Planck:2020olo} (30--545\,GHz) data to produce a
minimum-variance CMB map over approximately
$13{,}000\;\mathrm{deg}^2$.  The map is convolved to a common
Gaussian beam with $\mathrm{FWHM} = 1.6'$ and has a mean white
noise depth of approximately $15\;\mu\mathrm{K}$-arcmin.

We also use single-frequency ACT DR6 co-added, source-free temperature maps~\cite{Naess2025} at central frequencies of 98\,GHz, 150\,GHz, and 220\,GHz.  The 98\,GHz channel is referred to as ``90\,GHz'' throughout, following the convention of the ACT data products.  We use the night-only co-adds, which have reduced beam systematics on small scales. As a robustness check, we also repeat the analysis using the \texttt{daynight} co-adds, which combine daytime and nighttime observations and have somewhat smaller noise (Section~\ref{sec:map_robustness}).

\paragraph{Masking and pixel weights.}
We apply two layers of masking.  First, we apply the Planck
\texttt{GAL070} galactic foreground mask, which retains 70\% of the
sky.  Second, we define the CMB pixel weight as
\begin{equation}\label{eq:wcmb}
    W_{\rm CMB}(\btheta) =
    \begin{cases}
    1\,, & \text{if the noise in pixel } \btheta \text{ is}
    \le 70\;\mu\text{K-arcmin for both 90 and 150\,GHz}\,,\\
    0\,, & \text{otherwise}\,.
    \end{cases}
\end{equation}
The noise threshold is applied independently at each frequency, and
we take the intersection so that both channels share the same sky
footprint.  No tSZ cluster mask is applied; we have verified that
including one does not change our results.

\paragraph{Beam equalization.}
We match the effective beams of the 150 and 220\,GHz maps to that
of the 90\,GHz channel by multiplying in harmonic space:
\begin{equation}\label{eq:beam_eq}
    T^{\nu}_{\ell m}\big|_{b_{90}} =
    T^{\nu}_{\ell m}\,\frac{b_\ell^{90}}{b_\ell^{\nu}}\,,
\end{equation}
where $b_\ell^\nu$ denotes the beam transfer function at frequency
$\nu$.  After this operation, all single-frequency channels share the same effective beam $b_\ell^{90}$, which is the beam that appears in the theory prediction (Eq.~\ref{eq:ClTG}); for NILC, the corresponding beam is $b_\ell^{N}$ (defined below). 

\paragraph{Frequency differences.}
From the beam-equalized single-frequency maps, we construct three
frequency-difference maps for use as null tests:
\begin{itemize}
\item \emph{150$-$90 difference:} $T^{150-90}_{\ell m} =
    T^{150}_{\ell m}\big|_{b_{90}} - T^{90}_{\ell m}$, which
    cancels signals with a blackbody spectrum and serves as a null
    test for frequency-dependent systematics.
\item \emph{220$-$150 difference:} $T^{220-150}_{\ell m} =
    T^{220}_{\ell m}\big|_{b_{90}} - T^{150}_{\ell m}\big|_{b_{90}}$.
\item \emph{220$-$90 difference:} $T^{220-90}_{\ell m} =
    T^{220}_{\ell m}\big|_{b_{90}} - T^{90}_{\ell m}$.
\end{itemize}

\paragraph{220\,GHz noise mask.}
As a robustness check, we define a stricter masking condition that
additionally requires the 220\,GHz white-noise level to be below
$300\;\mu\mathrm{K}$-arcmin in each pixel.  This mask excludes the
noisiest 220\,GHz regions and slightly reduces the effective
sky footprint.  All maps and galaxy catalogs are re-processed with
this mask applied; the results are described in
Section~\ref{sec:map_robustness}.

We compute cross-spectra
$C_\ell^{TG} = \langle {X^G_{\ell m}}\,T^*_{\ell m}\rangle/(2\ell+1)$
for each CMB map (NILC, single-frequency, and frequency-difference).

%%%%%%%%%%%%%%%%%%%%%%%%%%%%%%%%%%%%%%%%%%%%%%%%%%%%%%%%%%%%%%%%%%%%%%%%%%
\section{Results}\label{sec:results}
%%%%%%%%%%%%%%%%%%%%%%%%%%%%%%%%%%%%%%%%%%%%%%%%%%%%%%%%%%%%%%%%%%%%%%%%%%

\subsection{Velocity reconstruction}\label{sec:velrecon}

We apply the pipeline of Section~\ref{sec:pipeline} to the data
described in Section~\ref{sec:data}.  The galaxy density field
$\rho_g(\bx)$ (Eq.~\ref{eq:rho_g}) is computed on a three-dimensional
grid with pixel size 10\,Mpc and a bounding-box padding of 3000\,Mpc
to suppress periodic boundary artifacts.  The Wiener filter
(Eq.~\ref{eq:U_filter}) is applied with a cutoff
$k_{\mathrm{max}} = 0.05\;\mathrm{Mpc}^{-1}$ and a fiducial galaxy
bias $b_g = 2.054$.

The resulting transverse velocity components
$\hat{v}_\theta(\bx)$ and $\hat{v}_\phi(\bx)$ are evaluated at each
galaxy position and painted onto HEALPix maps at $N_{\rm side} = 4096$.\footnote{We reconstruct velocities on HEALPix maps and compute alms via healpy. We keep the CMB maps in their pixell CAR pixelization and compute alms via pixell.curvedsky. We reproject the Planck galactic-plane mask from galactic onto the pixell CAR grid before multiplying it with the CMB maps; both HEALPix and pixell maps are in equatorial coordinates.}
We construct two sets of maps: \emph{galaxy-only} maps, which use only
the galaxy sum in Eq.~\eqref{eq:Xa_def}, and \emph{random-subtracted}
maps, which include the $(N_g/N_r)\sum_\beta$ random-catalog term from
Eq.~\eqref{eq:rho_g}.  The random subtraction removes the imprint of
the survey selection function while preserving the correlated velocity
signal.

\begin{figure*}[t]
    \centering
    \includegraphics[width=0.85\textwidth]{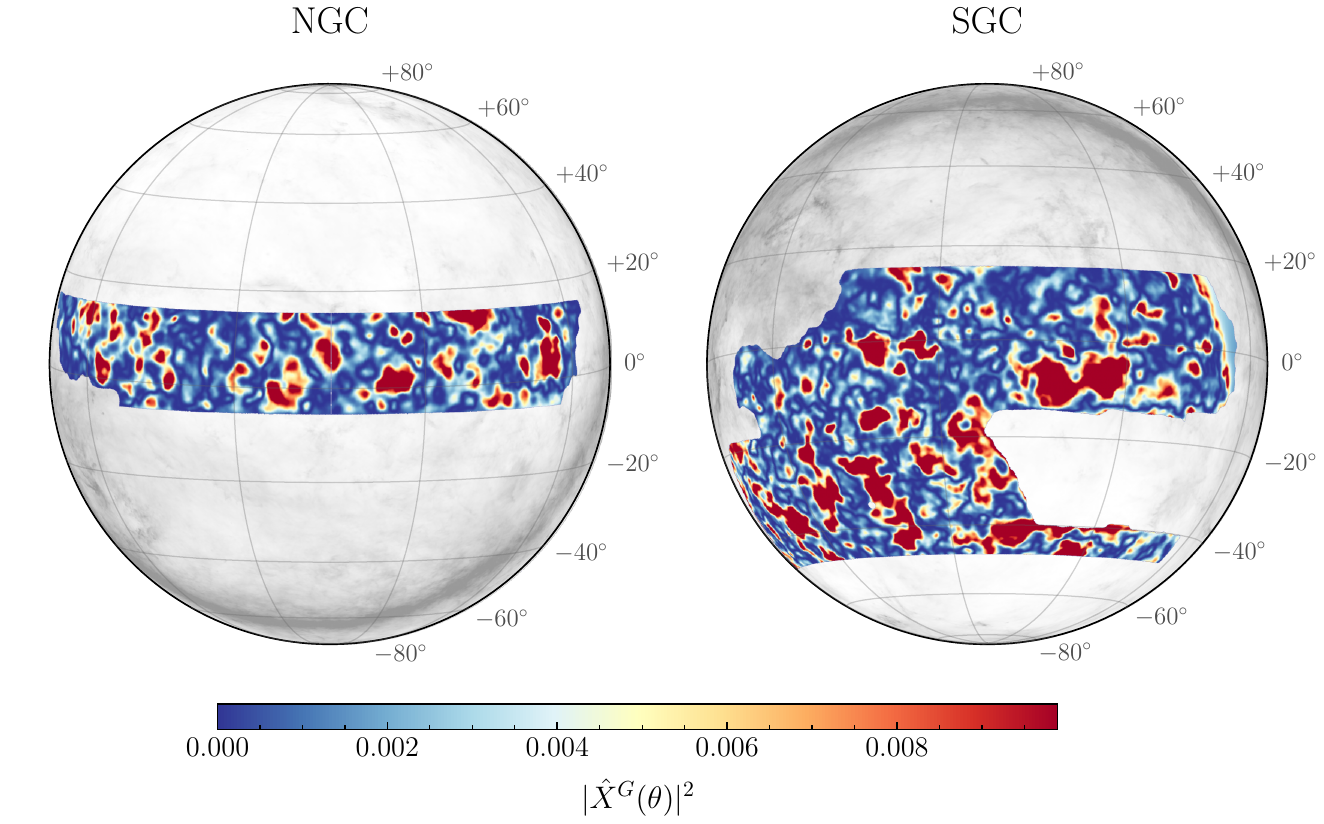}
    \caption{Orthographic projection of $|X^G(\hat n)|^2$ for the DESI-LS main LRG $\times$ ACT DR6 dataset (random-subtracted). Left: NGC. Right: SGC. The colored region shows $|X^G(\hat n)|^2$ at full opacity within the survey footprint (DESI-LS $\cap$ ACT noise mask); the grayscale background shows the Planck 353\,GHz GNILC thermal dust emission~\cite{PlanckDust2016} at reduced opacity. $X^G(\hat n)$ is the inverse SHT of $X^G_{\ell m}$ truncated at $\ell_{\max}=2000$ and smoothed with a $50'$ FWHM Gaussian beam; the monopole and dipole have been removed within the footprint before squaring. RA/Dec coordinate lines are overlaid at $40^\circ$ and $20^\circ$ intervals, respectively.}
    \label{fig:vgmap}
\end{figure*}

Figure~\ref{fig:vgmap} shows $|X^G(\hat n)|^2$, the squared real-space gradient mode of the random-subtracted velocity field, in orthographic projection centered on the northern (NGC, left) and southern (SGC, right) Galactic caps. We obtain $X^G(\hat n)$ as the inverse spherical-harmonic transform of $X^G_{\ell m}$ truncated at $\ell_{\max}=2000$ and smoothed with a $50'$ FWHM Gaussian beam; the monopole and dipole are removed within the survey footprint before squaring. The colored region shows the survey footprint (DESI-LS $\cap$ ACT noise mask); the grayscale background shows the Planck 353\,GHz GNILC dust emission~\cite{PlanckDust2016} in the complementary sky area.

\subsection{Signal measurements} \label{sec:signal_meas}

We begin with the foreground-cleaned NILC map, which provides the lowest-noise constraints on $b_{\rm ML}$.  The NILC map minimizes the total power of foregrounds and noise using all available frequency channels~\cite{ACT:2023wcq}, making it the minimum-variance choice for signals with a blackbody frequency dependence, such as the ML effect.

\paragraph{NILC beam.} The NILC map is reconvolved to a Gaussian beam $b_\ell^{N}$ with $\mathrm{FWHM} = 1.6'$. 
Rather than correcting the data to $b_\ell^{90}$, we convolve the theory template (Eq.~\ref{eq:ClTG}) with $b_\ell^{N}$ directly when fitting the NILC bandpowers.

For the combined NGC+SGC footprint, the extended sample yields
\begin{equation}
    b_{\rm ML} = 1.24 \pm 0.26 \quad (4.8\sigma)\,,
\end{equation}
consistent with the fiducial halo-model prediction. 
The main sample gives $b_{\rm ML} = 0.93 \pm 0.25$ ($3.7\sigma$).

We also fit $b_{\rm ML}$ independently to each single-frequency map (90, 150, and 220\,GHz).  For the extended sample (NGC+SGC) we find:
\begin{itemize}
\item 90\,GHz: $b_{\rm ML} = 0.93 \pm 0.36$ ($2.6\sigma$, $\chi^2\!/\mathrm{dof} = 0.96$,
  $\mathrm{PTE(gof)} = 0.44$),
\item 150\,GHz: $b_{\rm ML} = 1.14 \pm 0.32$ ($3.6\sigma$, $\chi^2\!/\mathrm{dof} = 0.47$,
  $\mathrm{PTE(gof)} = 0.80$),
\item 220\,GHz: $b_{\rm ML} = 3.08 \pm 1.29$ ($2.4\sigma$, $\chi^2\!/\mathrm{dof} = 0.47$,
  $\mathrm{PTE(gof)} = 0.80$).
\end{itemize}
where we also report the $\chi^2$ of the one-parameter $b_{\rm ML}$-fit, and the associated ``goodness of fit'' $p$-value, denoted PTE(gof).
The main sample gives consistent amplitudes for 90 and 150~GHz: $b_{\rm ML} = 0.74 \pm 0.35$ (90\,GHz), $0.62 \pm 0.31$ (150\,GHz); whereas for 220~GHz, $b_{\rm ML}=2.92 \pm 1.28$.
Table~\ref{tab:signal} collects the full set of results for both samples and all footprints.

\begin{figure*}[t!]
    \centering
    \includegraphics[width=0.9\textwidth]{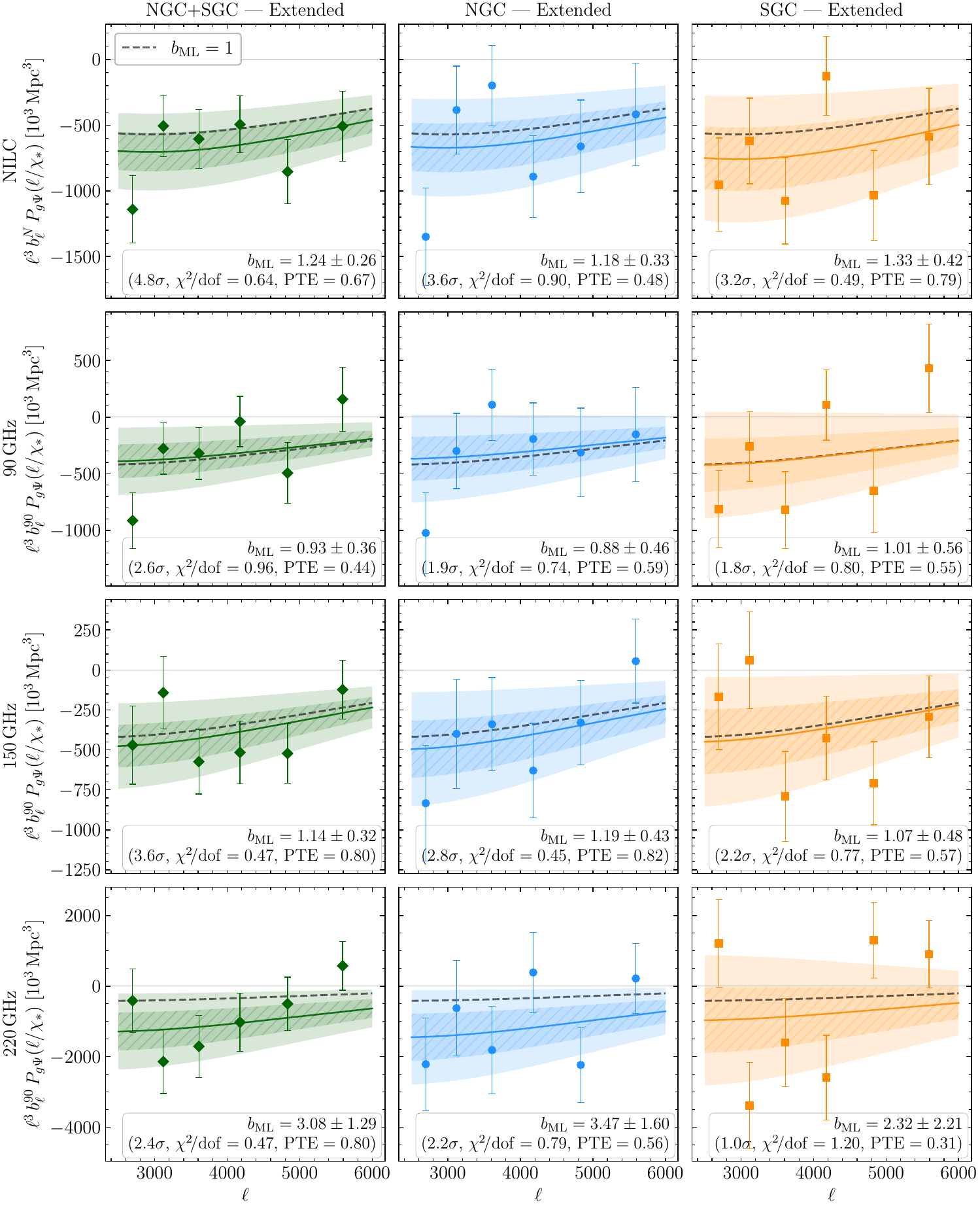}
    \caption{The NILC and 90, 150 and 220 GHz bandpowers for the DESI-LS extended LRG sample. Panels show the gradient mode $X^G_{\ell m}$. From left to right: NGC+SGC (inverse-variance weighted), NGC, SGC. Solid curves: best-fit $b_{\rm ML}$ template; darker hatched bands: $1\sigma$; lighter smooth bands: $2\sigma$. Dashed curve corresponds to $b_{\rm ML} = 1$.}
    \label{fig:vg_all}
\end{figure*}

Figure~\ref{fig:vg_all} shows the gradient-mode bandpowers for all four CMB maps (NILC, 90, 150, and 220\,GHz) and all three footprints (NGC+SGC, NGC, SGC) for the extended sample.  The NILC results (first row) follow the ML template across the full multipole range. 
The NILC map, which minimizes the combined foreground and noise power across all channels, recovers an amplitude consistent with unity ($b_{\rm ML} = 1.24 \pm 0.26$).

\paragraph{Sub-region (NGC vs SGC) analysis.} For the extended sample, the NGC and SGC sub-regions yield $b_{\rm ML} = 1.18 \pm 0.33$ ($3.6\sigma$) and $1.33 \pm 0.42$ ($3.2\sigma$), respectively. 
The main sample shows consistent amplitudes across sub-regions: $b_{\rm ML} = 0.94 \pm 0.36$ (NGC) and $0.93 \pm 0.36$ (SGC).

The individual-frequency bandpowers are shown alongside the NILC results in Figure~\ref{fig:vg_all} (rows 2--4).
At 150\,GHz (third row), the bandpowers have a high goodness of fit ($\mathrm{PTE(gof)} = 0.80$); the NGC and SGC amplitudes are $1.19 \pm 0.43$ and $1.07 \pm 0.48$, both consistent with unity.
At 220\,GHz (fourth row), the larger scatter reflects the higher noise level, but the template shape remains an acceptable fit.
The NGC and SGC amplitudes are consistent across samples (extended $3.47 \pm 1.60$ NGC vs $2.32 \pm 2.21$ SGC, while for the main, $3.66 \pm 1.89$ NGC vs $2.29 \pm 1.74$ SGC).

\begin{table*}[t]
\centering

\caption{Gradient-mode ({$X_{\ell m}^G$}) $b_{\rm ML}$ fits for all signal maps and footprints, over $2500 \le \ell \le 6000$ (6 log-spaced bins). $\mathrm{PTE(gof)}$: goodness-of-fit probability at best-fit $b_{\rm ML}$. NGC+SGC values are obtained by inverse-variance weighting of per-cap bandpowers.}
\label{tab:signal}
\begin{tabular}{llcccc@{\hspace{18pt}}cccc}
\hline\hline
 & & \multicolumn{4}{c}{Main} & \multicolumn{4}{c}{Extended} \\
Map & Footprint & $b_{\rm ML} \pm \sigma$ & SNR & $\chi^2\!/\mathrm{dof}$ & $\mathrm{PTE(gof)}$
              & $b_{\rm ML} \pm \sigma$ & SNR & $\chi^2\!/\mathrm{dof}$ & $\mathrm{PTE(gof)}$ \\
\hline\hline
\multicolumn{10}{l}{\emph{NILC}} \\
 & NGC+SGC  & $0.93 \pm 0.25$ & $3.7$ & $1.03$ & $0.40$
           & $1.24 \pm 0.26$ & $4.8$ & $0.64$ & $0.67$ \\
 & NGC      & $0.94 \pm 0.36$ & $2.6$ & $1.19$ & $0.31$
           & $1.18 \pm 0.33$ & $3.6$ & $0.90$ & $0.48$ \\
 & SGC      & $0.93 \pm 0.36$ & $2.6$ & $1.21$ & $0.30$
           & $1.33 \pm 0.42$ & $3.2$ & $0.49$ & $0.79$ \\
\hline
\multicolumn{10}{l}{\emph{90\,GHz}} \\
 & NGC+SGC  & $0.74 \pm 0.35$ & $2.1$ & $0.68$ & $0.64$
           & $0.93 \pm 0.36$ & $2.6$ & $0.96$ & $0.44$ \\
 & NGC      & $0.75 \pm 0.50$ & $1.5$ & $1.08$ & $0.37$
           & $0.88 \pm 0.46$ & $1.9$ & $0.74$ & $0.59$ \\
 & SGC      & $0.75 \pm 0.49$ & $1.5$ & $0.71$ & $0.62$
           & $1.01 \pm 0.56$ & $1.8$ & $0.80$ & $0.55$ \\[3pt]
\multicolumn{10}{l}{\emph{150\,GHz}} \\
 & NGC+SGC  & $0.62 \pm 0.31$ & $2.0$ & $0.73$ & $0.60$
           & $1.14 \pm 0.32$ & $3.6$ & $0.47$ & $0.80$ \\
 & NGC      & $0.85 \pm 0.48$ & $1.8$ & $0.50$ & $0.78$
           & $1.19 \pm 0.43$ & $2.8$ & $0.45$ & $0.82$ \\
 & SGC      & $0.46 \pm 0.42$ & $1.1$ & $1.18$ & $0.31$
           & $1.07 \pm 0.48$ & $2.2$ & $0.77$ & $0.57$ \\[3pt]
\multicolumn{10}{l}{\emph{220\,GHz}} \\
 & NGC+SGC  & $2.92 \pm 1.28$ & $2.3$ & $0.46$ & $0.80$
           & $3.08 \pm 1.29$ & $2.4$ & $0.47$ & $0.80$ \\
 & NGC      & $3.66 \pm 1.89$ & $1.9$ & $1.11$ & $0.35$
           & $3.47 \pm 1.60$ & $2.2$ & $0.79$ & $0.56$ \\
 & SGC      & $2.29 \pm 1.74$ & $1.3$ & $0.99$ & $0.42$
           & $2.32 \pm 2.21$ & $1.0$ & $1.20$ & $0.31$ \\
\hline\hline
\end{tabular}
\end{table*}

\begin{table*}[t]
\centering
\caption{Null-test $b_{\rm ML}$ fits over $2500 \le \ell \le 6000$ (6 log-spaced bins).  Curl-mode ({$X_{\ell m}^C$}) cross-spectra test for systematic contamination in the spin-1 decomposition; frequency-difference ({$X_{\ell m}^G$}) cross-spectra test for departures from blackbody frequency dependence; NGC$-$SGC ({$X_{\ell m}^G$}) tests consistency of the two galactic caps.  All should yield $b_{\rm ML} = 0$. $\mathrm{PTE(gof)}$: goodness-of-fit probability. $p_0$: probability of the data under the null hypothesis of no signal. NGC+SGC values are obtained by inverse-variance weighting of per-cap bandpowers.}
\label{tab:null}
\begin{tabular}{llccccc@{\hspace{20pt}}ccccc}
\hline\hline
 & & \multicolumn{5}{c}{Main} & \multicolumn{5}{c}{Extended} \\
Map & Footprint & $b_{\rm ML} \pm \sigma$ & $\mathrm{SNR}$ & $\chi^2\!/\mathrm{dof}$ & $\mathrm{PTE(gof)}$ & $p_0$
              & $b_{\rm ML} \pm \sigma$ & $\mathrm{SNR}$ & $\chi^2\!/\mathrm{dof}$ & $\mathrm{PTE(gof)}$ & $p_0$ \\
\hline\hline
\multicolumn{12}{l}{\emph{Curl mode ({$X_{\ell m}^C$})}} \\[4pt]
\multicolumn{12}{l}{\emph{NILC}} \\
 & NGC+SGC  & $+0.14 \pm 0.25$ & $0.5$ & $0.92$ & $0.47$ & $0.58$
           & $-0.11 \pm 0.25$ & $0.5$ & $0.83$ & $0.53$ & $0.65$ \\
 & NGC      & $-0.11 \pm 0.36$ & $0.3$ & $0.50$ & $0.78$ & $0.77$
           & $-0.04 \pm 0.33$ & $0.1$ & $0.50$ & $0.78$ & $0.90$ \\
 & SGC      & $+0.36 \pm 0.35$ & $1.0$ & $1.45$ & $0.20$ & $0.30$
           & $-0.20 \pm 0.38$ & $0.5$ & $0.76$ & $0.58$ & $0.59$ \\
\hline
\multicolumn{12}{l}{\emph{90\,GHz}} \\
 & NGC+SGC  & $+0.41 \pm 0.35$ & $1.2$ & $0.67$ & $0.64$ & $0.24$
           & $+0.29 \pm 0.37$ & $0.8$ & $0.83$ & $0.53$ & $0.43$ \\
 & NGC      & $+0.58 \pm 0.50$ & $1.2$ & $0.42$ & $0.83$ & $0.25$
           & $+0.82 \pm 0.48$ & $1.7$ & $0.13$ & $0.98$ & $0.09$ \\
 & SGC      & $+0.24 \pm 0.48$ & $0.5$ & $0.81$ & $0.54$ & $0.61$
           & $-0.43 \pm 0.57$ & $0.8$ & $1.03$ & $0.40$ & $0.45$ \\[3pt]
\multicolumn{12}{l}{\emph{150\,GHz}} \\
 & NGC+SGC  & $+0.02 \pm 0.30$ & $0.1$ & $0.82$ & $0.54$ & $0.95$
           & $-0.31 \pm 0.31$ & $1.0$ & $1.01$ & $0.41$ & $0.31$ \\
 & NGC      & $-0.38 \pm 0.41$ & $0.9$ & $0.62$ & $0.68$ & $0.34$
           & $-0.41 \pm 0.39$ & $1.0$ & $1.07$ & $0.37$ & $0.30$ \\
 & SGC      & $+0.48 \pm 0.44$ & $1.1$ & $1.33$ & $0.25$ & $0.27$
           & $-0.17 \pm 0.49$ & $0.3$ & $0.24$ & $0.95$ & $0.74$ \\[3pt]
\multicolumn{12}{l}{\emph{220\,GHz}} \\
 & NGC+SGC  & $+0.75 \pm 1.18$ & $0.6$ & $1.52$ & $0.18$ & $0.52$
           & $+0.14 \pm 1.25$ & $0.1$ & $0.93$ & $0.46$ & $0.91$ \\
 & NGC      & $+0.66 \pm 1.66$ & $0.4$ & $0.53$ & $0.75$ & $0.69$
           & $+0.42 \pm 1.61$ & $0.3$ & $1.08$ & $0.37$ & $0.79$ \\
 & SGC      & $+0.82 \pm 1.67$ & $0.5$ & $1.64$ & $0.14$ & $0.62$
           & $-0.30 \pm 1.98$ & $0.2$ & $1.60$ & $0.16$ & $0.88$ \\
\hline\hline
\multicolumn{12}{l}{\emph{Frequency differences ({$X_{\ell m}^G$})}} \\[4pt]
\multicolumn{12}{l}{\emph{$150{-}90$}} \\
 & NGC+SGC  & $-0.19 \pm 0.38$ & $0.5$ & $0.48$ & $0.79$ & $0.62$
           & $+0.12 \pm 0.39$ & $0.3$ & $0.94$ & $0.46$ & $0.76$ \\
 & NGC      & $+0.14 \pm 0.54$ & $0.3$ & $0.63$ & $0.68$ & $0.79$
           & $+0.44 \pm 0.50$ & $0.9$ & $0.43$ & $0.83$ & $0.38$ \\
 & SGC      & $-0.54 \pm 0.55$ & $1.0$ & $0.24$ & $0.94$ & $0.32$
           & $-0.37 \pm 0.62$ & $0.6$ & $0.81$ & $0.54$ & $0.55$ \\[3pt]
\hline
\multicolumn{12}{l}{\emph{$220{-}90$}} \\
 & NGC+SGC  & $+2.02 \pm 1.29$ & $1.6$ & $0.48$ & $0.79$ & $0.12$
           & $+2.17 \pm 1.30$ & $1.7$ & $0.44$ & $0.82$ & $0.09$ \\
 & NGC      & $+2.77 \pm 1.84$ & $1.5$ & $1.02$ & $0.40$ & $0.13$
           & $+2.56 \pm 1.59$ & $1.6$ & $0.73$ & $0.60$ & $0.11$ \\
 & SGC      & $+1.31 \pm 1.82$ & $0.7$ & $0.88$ & $0.49$ & $0.47$
           & $+1.40 \pm 2.27$ & $0.6$ & $1.33$ & $0.25$ & $0.54$ \\[3pt]
\hline
\multicolumn{12}{l}{\emph{$220{-}150$}} \\
 & NGC+SGC  & $+2.28 \pm 1.25$ & $1.8$ & $0.43$ & $0.83$ & $0.07$
           & $+2.02 \pm 1.26$ & $1.6$ & $0.45$ & $0.81$ & $0.11$ \\
 & NGC      & $+2.86 \pm 1.78$ & $1.6$ & $1.17$ & $0.32$ & $0.11$
           & $+2.34 \pm 1.53$ & $1.5$ & $0.78$ & $0.56$ & $0.13$ \\
 & SGC      & $+1.73 \pm 1.76$ & $1.0$ & $0.88$ & $0.49$ & $0.33$
           & $+1.36 \pm 2.22$ & $0.6$ & $1.40$ & $0.22$ & $0.54$ \\
\hline\hline
\multicolumn{12}{l}{\emph{NGC$-$SGC ({$X_{\ell m}^G$})}} \\[4pt]
% \multicolumn{12}{l}{\emph{NILC}} \\
 & NILC & $+0.03 \pm 0.51$ & $0.1$ & $1.39$ & $0.22$ & $0.95$
           & $-0.06 \pm 0.54$ & $0.1$ & $0.77$ & $0.57$ & $0.92$ \\
% \multicolumn{12}{l}{\emph{90\,GHz}} \\
 & 90\,GHz & $+0.02 \pm 0.71$ & $0.0$ & $1.11$ & $0.35$ & $0.98$
           & $-0.06 \pm 0.73$ & $0.1$ & $0.60$ & $0.70$ & $0.94$ \\
% \multicolumn{12}{l}{\emph{150\,GHz}} \\
 & 150\,GHz & $+0.30 \pm 0.63$ & $0.5$ & $0.99$ & $0.42$ & $0.63$
           & $+0.06 \pm 0.65$ & $0.1$ & $0.75$ & $0.59$ & $0.92$ \\
% \multicolumn{12}{l}{\emph{220\,GHz}} \\
 & 220\,GHz & $+1.16 \pm 2.61$ & $0.4$ & $1.64$ & $0.15$ & $0.66$
           & $+1.33 \pm 2.73$ & $0.5$ & $1.50$ & $0.19$ & $0.63$ \\
\hline\hline
\end{tabular}
\end{table*}

\subsection{Null tests}\label{sec:nulltests}

\begin{figure*}[t!]
    \centering
    \includegraphics[width=0.9\textwidth]{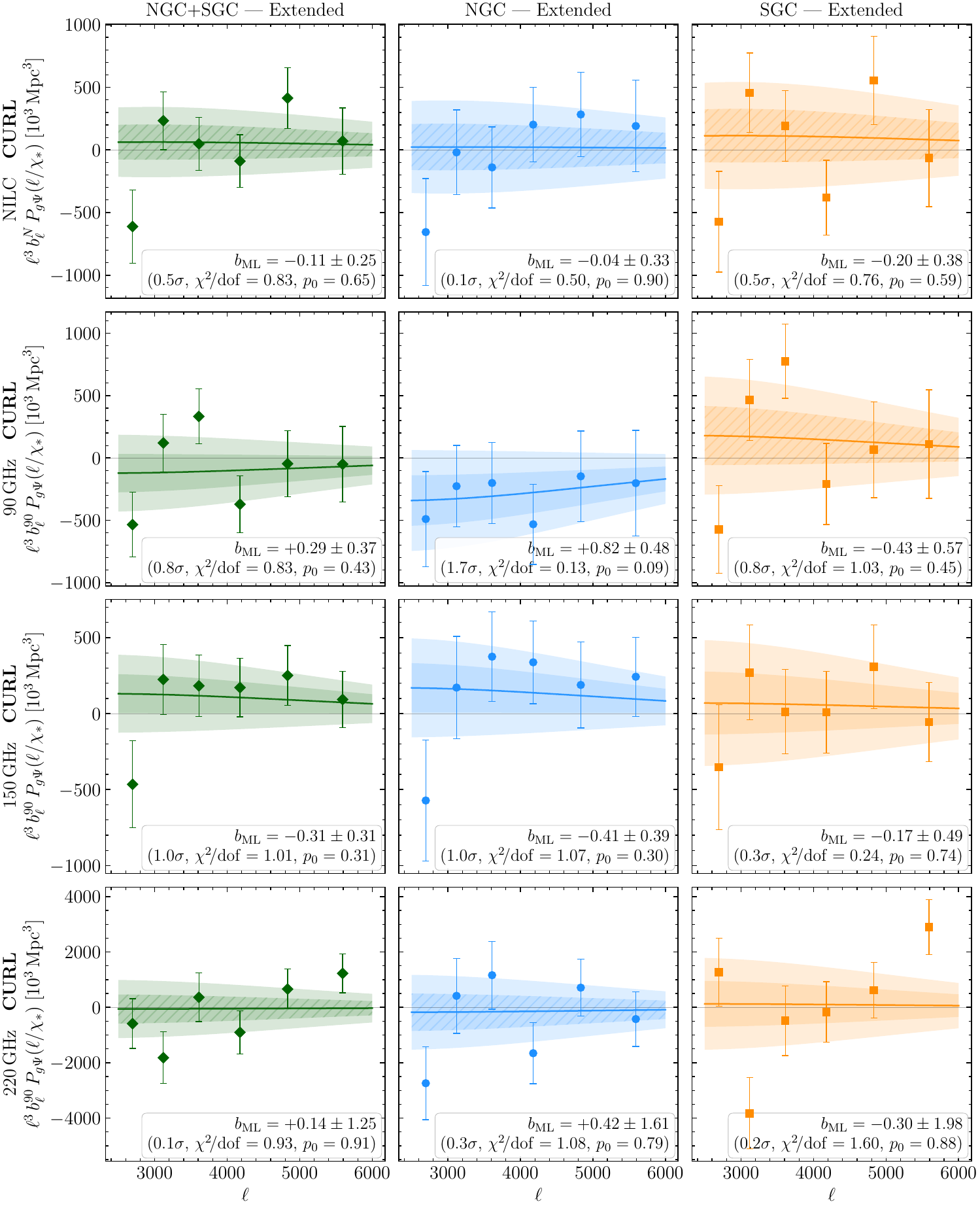}
    \caption{The NILC and 90, 150 and 220 GHz bandpowers for the DESI-LS extended LRG sample. Panels show the \emph{curl} mode $X^C_{\ell m}$. From left to right: NGC+SGC (inverse-variance weighted), NGC, SGC. Solid curves: best-fit $b_{\rm ML}$ template; darker hatched bands: $1\sigma$; lighter smooth bands: $2\sigma$.}
    \label{fig:null_vc}
\end{figure*}

\begin{figure*}[t!]
    \centering
    \includegraphics[width=0.9\textwidth]{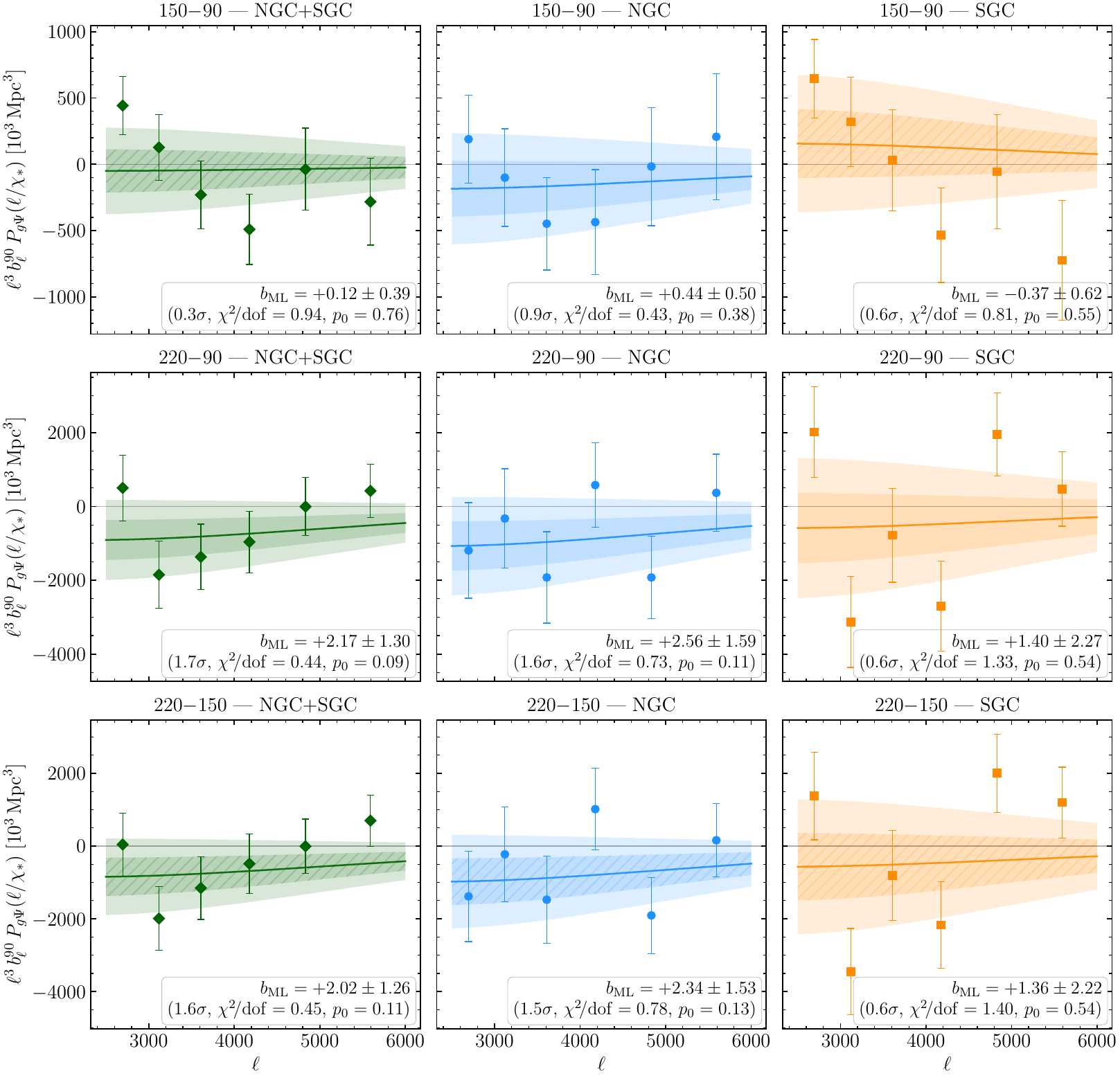}
    \caption{Gradient-mode null-test bandpowers from frequency-difference maps for the DESI-LS extended LRG sample. \emph{Rows from top to bottom:} $150{-}90$, $220{-}90$, $220{-}150$. \emph{Columns from left to right:} NGC+SGC combined, NGC, SGC. The $150{-}90$ difference is consistent with zero for all footprints, as expected for a blackbody signal. Markers and bands as in  Figures~\ref{fig:vg_all}~and~\ref{fig:null_vc}.}
    \label{fig:null_diff_vg}
\end{figure*}

We perform three classes of null tests: curl mode, frequency differences, and NGC-SGC consistency, described below.
All null tests are collected in Table~\ref{tab:null}.

\paragraph{Curl mode.}
As shown in Eq.\ (\ref{eq:cltg_cltc_flat}), the curl-mode correlation
$C_\ell^{TC}$ vanishes by parity symmetry; non-zero power would indicate
parity-asymmetric systematics.
The curl-mode cross-spectra for all four CMB maps are shown
in Figure~\ref{fig:null_vc}.
For the extended sample (NGC+SGC), all four channels yield $b_{\rm ML}$ consistent with zero at low significance:
NILC ($0.5\sigma$, $p_0 = 0.65$),
90\,GHz ($0.8\sigma$, $p_0 = 0.43$),
150\,GHz ($1.0\sigma$, $p_0 = 0.31$), and
220\,GHz ($0.1\sigma$, $p_0 = 0.91$).
The $\chi^2\!/\mathrm{dof}$ values range from $0.13$ to $1.60$, with goodness-of-fit probabilities in $[0.16, 0.98]$ -- all above $0.05$. The NGC and SGC sub-regions are individually consistent with zero for all channels (Table~\ref{tab:null}).

For the main sample (NGC+SGC), the curl-mode amplitudes are again consistent with zero:
NILC ($0.5\sigma$, $p_0 = 0.58$),
90\,GHz ($1.2\sigma$, $p_0 = 0.24$),
150\,GHz ($0.1\sigma$, $p_0 = 0.95$), and
220\,GHz ($0.6\sigma$, $p_0 = 0.52$).
No curl-mode test reaches $2\sigma$ significance for either sample. 

\paragraph{Frequency differences.}
Frequency-difference maps
cancel any signal with a blackbody spectrum, so a non-zero
amplitude in $150{-}90$, $220{-}90$, or $220{-}150$ would indicate
frequency-dependent contamination. 
The gradient-mode $X_{\ell m}^G$ bandpowers for all three frequency differences are shown in Figure~\ref{fig:null_diff_vg}.
The $150{-}90$ difference is the cleanest blackbody null: for the main sample (NGC+SGC), $b_{\rm ML} = -0.19 \pm 0.38$ ($p_0 = 0.62$, $\chi^2\!/\mathrm{dof} = 0.48$); for the extended sample, $b_{\rm ML} = +0.12 \pm 0.39$ ($p_0 = 0.76$, $\chi^2\!/\mathrm{dof} = 0.94$).
The NGC and SGC sub-regions are individually consistent with zero for both samples.

The $220{-}90$ and $220{-}150$ differences show positive amplitudes at $1.6$--$1.8\sigma$ in the combined footprint.
For the extended sample (NGC+SGC), $220{-}90$ gives $b_{\rm ML} = +2.17 \pm 1.30$ ($1.7\sigma$, $p_0 = 0.09$) and $220{-}150$ gives $+2.02 \pm 1.26$ ($1.6\sigma$, $p_0 = 0.11$).
For the main sample, $220{-}90$ gives $+2.02 \pm 1.29$ ($1.6\sigma$, $p_0 = 0.12$) and $220{-}150$ gives $+2.28 \pm 1.25$ ($1.8\sigma$, $p_0 = 0.07$).
All SGC frequency differences are consistent with zero ($p_0 > 0.32$ for both samples).

All frequency-difference null tests pass at 2$\sigma$ (i.e.\ $0.05 < p_0 < 0.95$).
Nevertheless, since the frequency-difference tests show the highest deviation from zero of any tests in our null test suite (up to 1.8$\sigma$), and CMB foregrounds have been highlighted as a concern for moving-lens pipelines~\cite{Hotinli:2023ywh, Hotinli:2024tjb, Beheshti:2024dxw}, we carry out additional foreground studies in \S\ref{sec:foreground}.

\paragraph{NGC--SGC footprint consistency.} For each CMB map, we also show in Table~\ref{tab:null} the difference between bandpowers of NGC and SGC as an additional systematics test; NILC gives $b_{\rm ML}^{\rm NGC-SGC} = -0.06 \pm 0.54$ ($p_0 = 0.92$) for the extended sample and $+0.03 \pm 0.51$ ($p_0 = 0.95$) for the main sample. The lowest $p_0$ across all eight NGC$-$SGC tests is $0.63$, well above the $0.05$ threshold.

\paragraph{Summary.}
Table~\ref{tab:null} collects the null-test results for both samples across all footprints.  
We perform 25 null tests per sample: 4 curl-mode channels and 3 frequency differences, each evaluated on 3 footprints (NGC+SGC, NGC, SGC), plus 4 NGC$-$SGC consistency tests (one per CMB map).
All null tests pass at $2\sigma$ ($p_0 > 0.05$).

\subsection{CMB map robustness checks}\label{sec:map_robustness}

We test the sensitivity of our results to the inclusion of daytime data in the CMB co-add, and the treatment of high-noise regions at 220\,GHz. 

\paragraph{Daynight maps.}
ACT DR6 also provides daytime-plus-night co-adds, which have somewhat smaller noise but larger beam systematics on small scales. We analyze these maps to check their consistency with the night-only baseline used in this work. Across our six single-frequency (NGC+SGC) signal channels (90, 150 and 220\,GHz, for the main and extended samples), shifts are typically of order $\mathcal{O}(0.1)$ sigma. No amplitude differs from the night-only baseline by more than $0.55\sigma$, with the largest shift at the 150\,GHz extended (NGC+SGC): $1.30 \pm 0.29$ (daynight) vs $1.14 \pm 0.32$ (night-only). While the daynight per-frequency error bars are $9$–$15\%$ smaller (the NILC maps, our tightest constraints, are produced from night-only maps), the daynight null tests show multiple main-sample deviations at $p_0 < 0.05$ including 220\,GHz and NGC-involving difference tests, all returning to $p_0 > 0.05$ in the night-only baseline. The extended sample shows no $p_0 < 0.05$ entries in either configuration.

\paragraph{220\,GHz noise mask.}
The 220\,GHz channel has the highest noise among the ACT frequency maps. We mask sky pixels where the 220\,GHz white-noise level exceeds $300\;\mu\mathrm{K}$-arcmin, reducing the effective footprint by $\sim\!10\%$. Across our six single-frequency (NGC+SGC) signal channels, amplitude shifts relative to the standard mask are small and mixed in sign, and the error bars are essentially unchanged ($\lesssim 10\%$ in either direction).

\section{Foreground studies}\label{sec:foreground}

In Table \ref{tab:null}, the largest null-test deviations involve 220 GHz data, with amplitudes of 0.6--1.8$\sigma$ across the $220{-}90$ and $220{-}150$ frequency-difference rows. Given the number of null tests in our test suite, this level of tension is consistent with being a statistical fluke, but it motivated us to carry out two additional foreground studies, in order to increase confidence that foreground contamination is not an issue.
The first test (\S\ref{ssec:planck_353_545}) uses Planck 353/545 GHz data and is empirical, and the second test (\S\ref{ssec:quijote_foregrounds}) is based on simulations.

\subsection{Constraints from Planck 353/545 GHz}
\label{ssec:planck_353_545}

\begin{figure*}[b]
    \centering
    \includegraphics[width=0.9\textwidth]{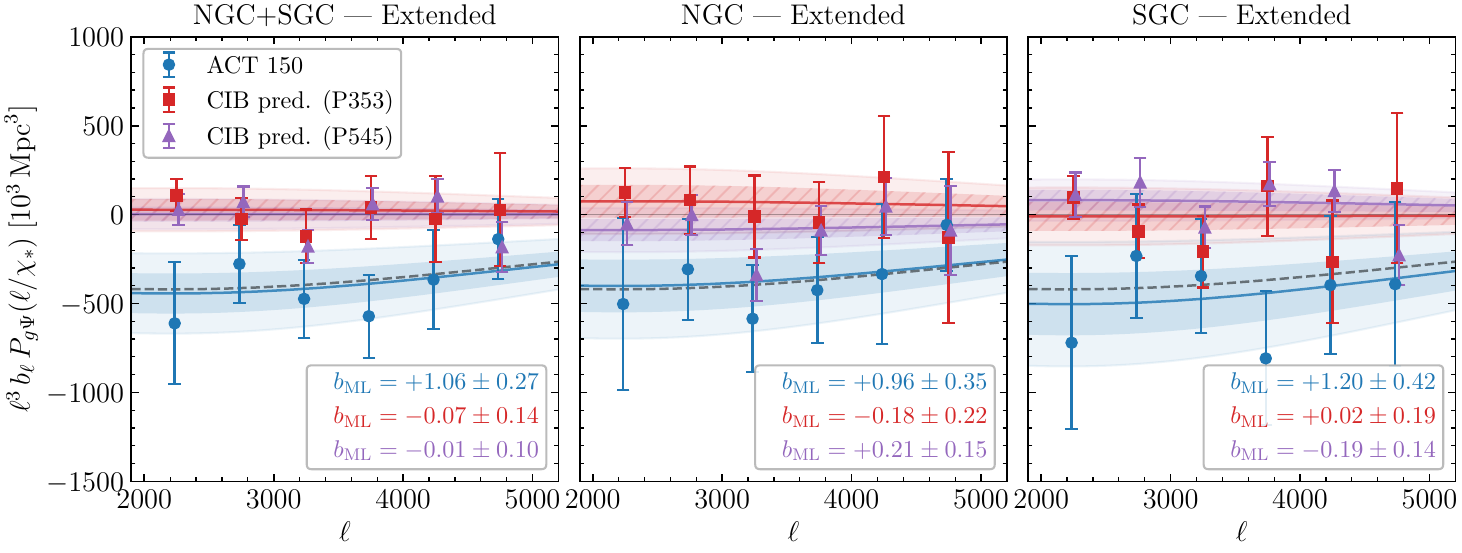}
    \caption{Planck NPIPE 353 and 545\,GHz cross-correlated with the velocity reconstruction, scaled to predict CIB     contamination at 150\,GHz (Eq.~\ref{eq:cib_scaling}).
    Blue: ACT 150\,GHz $\times\,X^G_{\ell m}$.
    Red/purple: Planck-derived CIB cross-spectra at 353/545\,GHz.
    Solid curves: best-fit $b_{\rm ML}$ template; hatched-darker/lighter bands: $1\sigma$/$2\sigma$.
    Dashed line corresponds to $b_{\rm ML} = 1$.
    From left to right: NGC+SGC, NGC, SGC.
    The analysis uses the GAL060 mask with $2^\circ$ apodization, $\Delta\ell = 500$, $2000 \le \ell \le 5000$, and CIB spectral parameters $\beta_d = 1.7$, $T_d = 10.7$\,K~\cite{ACT:2023wcq}.
    The legend in the first panel shows $b_{\rm ML}$ fits for the combined footprint; subsequent panels show per-cap values.}
    \label{fig:cib_planck}
\end{figure*}

The single-frequency amplitudes in Section~\ref{sec:signal_meas} show an elevated 220\,GHz amplitude ($b_{\rm ML} = 3.08 \pm 1.29$), and the frequency differences in Section~\ref{sec:nulltests} yield $1.6$--$1.8\sigma$ amplitudes for channels involving 220\,GHz.
To test whether this pattern reflects CIB contamination, we look at higher-frequency data.
Planck NPIPE~\cite{Planck:2020olo} provides full-sky temperature maps at frequencies up to 857\,GHz, well above the ACT frequency range (90--220\,GHz).
The CIB spectral response rises steeply with frequency, making the Planck 353 and 545\,GHz channels dominated by CIB contribution.

All results in this section use a more conservative angular mask, obtained from our default mask (see \S\ref{sec:data_cmb}) by using the Planck {\tt GAL060} mask instead of {\tt GAL070}. This more conservative mask mitigates Galactic dust contamination at 353/545 GHz. In particular, the ``ACT 150'' points in Figure \ref{fig:cib_planck} are slightly different from in the main paper, due to the different mask.

The Planck maps have higher noise and lower angular resolution than ACT.
Nevertheless, the CIB is so bright at 353/545 GHz that Planck data can be used to estimate the CIB contribution to the ACT moving-lens signal at lower frequencies.
We cross-correlate the NPIPE 353 and 545\,GHz intensity maps, reprojected to the ACT pixell geometry, with the gradient mode $X_{\ell m}^G$ of the moving-lens template, obtaining a cross power spectrum $C_\ell^{{\rm Planck} \times G}$.
To compare this power spectrum to our ACT measurements, we rescale to predict the CIB contribution at 150 GHz in ACT:
\begin{equation}\label{eq:cib_scaling}
    C_\ell^{({\rm CIB}\,150) \times G} =
    C_\ell^{({\rm Planck}\,\nu) \times G}
    \times \frac{f_{\rm CIB}(150)}{f_{\rm CIB}(\nu)}
    \times \frac{b_\ell^{90}}{b_\ell^{\nu}}\,,
\end{equation}
where $b_\ell^{90}$ is the ACT 90\,GHz beam transfer function (Section~\ref{sec:data_cmb}) {to which our ACT 150\,GHz map (and all single-frequency ACT maps) is convolved (Eq.~\ref{eq:beam_eq})}, and $b_\ell^{\nu}$ is the NPIPE beam transfer function at frequency~$\nu$. 
We model the CIB spectral response $f_{\rm CIB}(\nu)$ as a modified blackbody in CMB thermodynamic temperature units:
\begin{equation}\label{eq:f_cib}
    f_{\rm CIB}(\nu) \propto
    \frac{\nu^{\beta_d}\,B(\nu, T_d)}
    {(\partial B/\partial T)_{T_{\rm CMB}}}\,,
\end{equation}
with $\beta_d = 1.7$ and $T_d = 10.7$\,K following Ref.~\cite{ACT:2023wcq}.
The spectral ratios are $f_{\rm CIB}(150)/f_{\rm CIB}(353) = 0.054$ and $f_{\rm CIB}(150)/f_{\rm CIB}(545) = 0.0037$. 

To compute $C_\ell^{{\rm Planck} \times G}$, we run the pipeline described in
Section~\ref{sec:pipeline}, which consists of velocity reconstruction, surrogate-field normalization, and cross-spectrum computation; using the GAL060 galactic mask with $2^\circ$ apodization to reduce Galactic dust contamination at 353 and 545\,GHz.
All derived products are computed self-consistently for this footprint.
We bin in 6 bandpowers with $\Delta\ell = 500$ over $2000 \le \ell \le 5000$ and fit $b_{\rm ML}$ independently to the ACT 150\,GHz cross-spectrum and to each Planck-derived CIB cross-spectrum.
The upper multipole is lower than in the main analysis ($\ell_{\max} = 6000$) because the Planck beam suppresses signal at high~$\ell$ and these scales are dominated by noise in high-frequency maps.
As before, NGC and SGC are combined via inverse-variance weighting at the bandpower level.

In Figure \ref{fig:cib_planck}, we plot the quantity $C_\ell^{({\rm CIB}\,150) \times G}$ defined in Eq.\ (\ref{eq:cib_scaling}), converted to a measurement of $P_{g\Psi}(k)$ using Eq.\ (\ref{eq:ClTG}).
As explained above, this is a prediction for the CIB contribution to the ACT moving-lens signal at 150 GHz (assuming that 100\% of the signal in the Planck 353/545 maps is CIB).
Therefore, it can be compared directly to the ACT 150 GHz measurements (also shown in the plot).

The ACT 150\,GHz cross-spectrum gives $b_{\rm ML} = 1.06 \pm 0.27$ ($3.9\sigma$) for the combined footprint.
The Planck-derived CIB cross-spectra give $b_{\rm ML} = -0.07 \pm 0.14$ ($0.5\sigma$, 353\,GHz) and $-0.01 \pm 0.10$ ($0.1\sigma$, 545\,GHz), both consistent with zero.
NGC and SGC are individually consistent with zero: 
NGC gives $-0.18 \pm 0.22$ ($0.8\sigma$) and $+0.21 \pm 0.15$ ($1.4\sigma$); SGC gives $+0.02 \pm 0.19$ ($0.1\sigma$) and $-0.19 \pm 0.14$ ($1.4\sigma$).

Summarizing, when we use Planck 353/545 data to predict the CIB contamination to the ACT moving-lens signal, the results are consistent with zero contamination, with small error bars.
Using 545 GHz data, we predict CIB bias $\Delta b_{\rm ML} = 0.00 \pm 0.10$ (NGC+SGC extended) at 150 GHz, whereas the 150 GHz measured value is $b_{\rm ML} = 1.06 \pm 0.27$.
The CIB bias would be even smaller at 90 GHz.
This is strong empirical evidence that our moving lens measurement is not affected by CIB.

\subsection{Numerical results in a toy foreground model}
\label{ssec:quijote_foregrounds}

\begin{figure}
\centerline{\includegraphics[width=14cm]{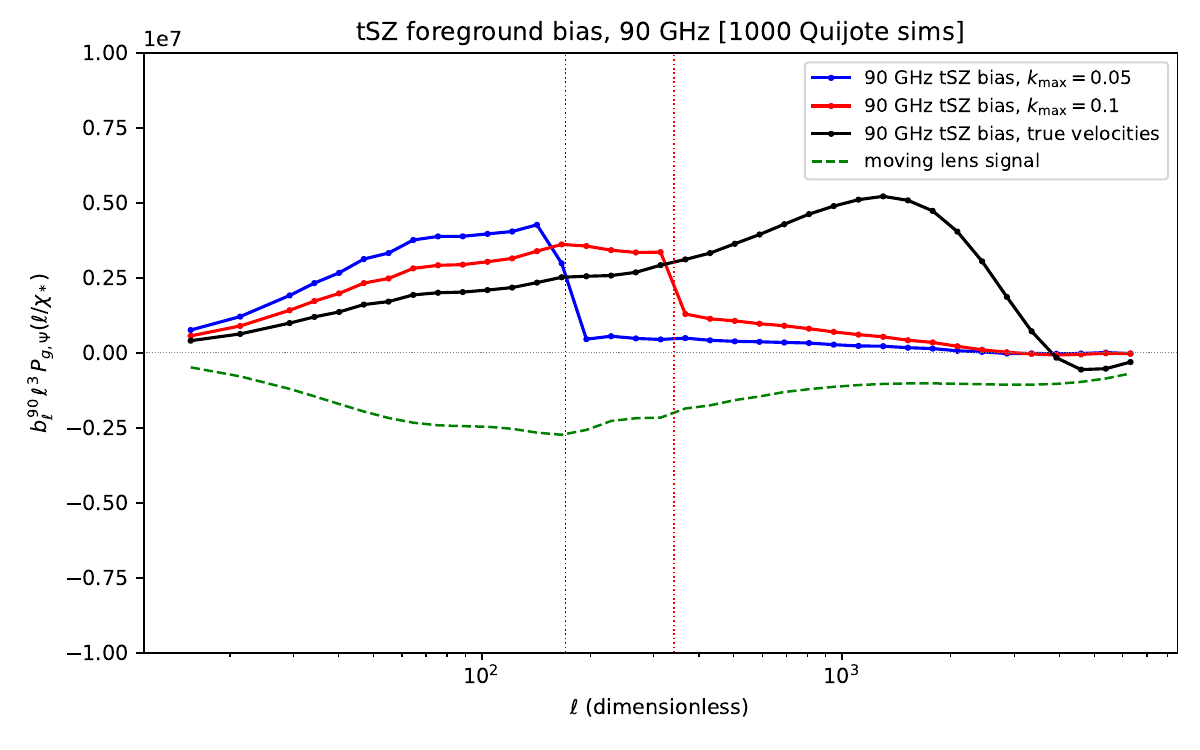}}
\centerline{\includegraphics[width=14cm]{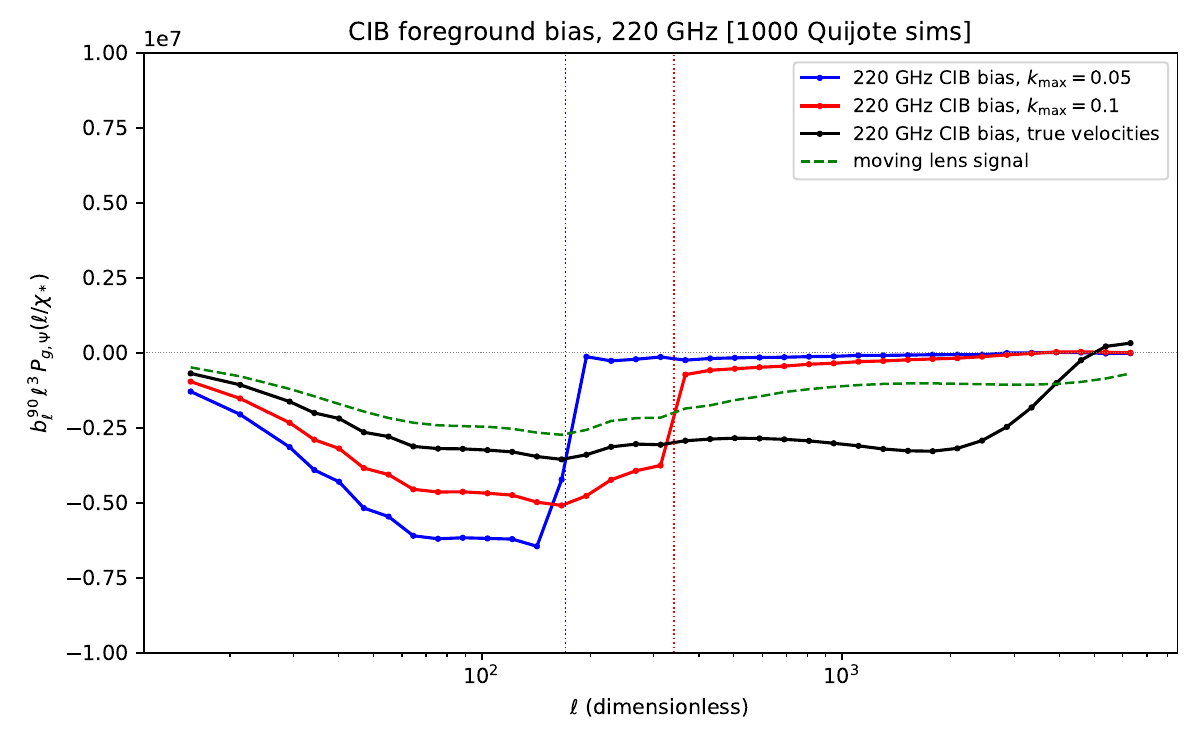}}
\caption{Foreground bias to $C_l^{TG}$ in a simplified Quijote-based pipeline at redshift $z_*=1$.
The top/bottom panels show the tSZ/CIB bias at the ACT frequency where the foreground is largest (90/220 GHz).
The blue curves are most representative of our DESILS pipeline, where the velocity reconstruction contains a sharp cutoff at $k_{\rm max}=0.05$ Mpc$^{-1}$ (see Eq.\ (\ref{eq:U_filter})).
We find that foreground biases are suppressed for $l \gtrsim k_{\rm max} \chi_* \approx 170$ (vertical lines), and smaller than the moving lens signal at high $l$. See text for more discussion.}
\label{fig:toy_moving_lens}
\end{figure}

To estimate the size of foreground biases to the moving-lens signal, we implemented a simplified pipeline based on Quijote simulations \cite{Villaescusa-Navarro:2019bje} and a toy foreground model.
This simplified pipeline neglects significant complications (curved-sky, lightcone evolution, more realistic foregrounds) but should correctly predict the approximate size of foreground biases.
We plan to do a more detailed study in future work.
The details of the pipeline and foreground model are presented in Appendix \ref{app:foreground_sims}.

Our main result from this pipeline is shown in Figure \ref{fig:toy_moving_lens}.
Recall that in our DESILS pipeline, the velocity reconstruction (see Eq.\ \ref{eq:U_filter}) contains a ``sharp'' cutoff at $k_{\rm max}=0.05$ Mpc$^{-1}$.
In cases with a sharp cutoff, we find that foreground biases from both tSZ and CIB are dramatically suppressed for $l \gtrsim (k_{\rm max} \chi_*)$ (blue/red curves in Figure\ \ref{fig:toy_moving_lens}), and are small compared to the moving-lens signal (dashed green curve).
(Note that the moving-lens signal in Figure \ref{fig:toy_moving_lens} differs from the fiducial model elsewhere in the paper, since it was computed using halo model parameters $(z,n_g^{3d})$ appropriate for Quijote.)

This may be surprising, since previous simulation-based studies have shown that correlations between extragalactic foreground emission gradients (tSZ and CIB) and the transverse velocity field can produce contamination an order of magnitude larger than the ML signal on single-frequency maps~\cite{Hotinli:2023ywh, Hotinli:2024tjb, Beheshti:2024dxw}. 
{Those analyses used true halo velocities from simulations; although they applied distortions to mimic reconstructed velocities~\citep[e.g.][]{Hotinli:2024tjb}, they did not impose a scale cut on the velocity modes.} By contrast, here we reconstruct velocities from the galaxy density field with a hard cutoff at $k_{\rm max} = 0.05\;\mathrm{Mpc}^{-1}$.
This cutoff produces a separation of scales: foreground contamination is significant at $\ell \lesssim k_{\rm max}\chi_* \approx 129$ but sharply drops to near zero at higher multipoles, well below our signal range ($\ell = 2500$--$6000$).
On the other hand, if true halo velocities are used in the pipeline (black curve in Figure\ \ref{fig:toy_moving_lens}), then foreground contamination extends to high $l$.
Intuitively, the small-scale physics of foreground emission is approximately uncorrelated with large-scale bulk motions, and therefore enforcing a clean separation of scales is sufficient to reduce the foregrounds to a negligible level.

Summarizing, we find that foreground bias to the moving-lens signal is predicted to be small in simulations, provided that two conditions are met:
\begin{enumerate}
\item The velocity reconstruction includes a ``hard'' cutoff at 3-d wavenumber $k_{\rm max}$, as in Eq.\ (\ref{eq:Udef_app}). Note that for realistic parameter values, the threshold wavenumber $l_*=k_{\rm max}\chi_*$ is much smaller than ``moving lens'' values ($l \gtrsim 2500$).
For example, if $k_{\rm max}=0.05$ and $z_*=0.7$, then $l_* = 129$.
\item The cross correlation with $T$ is either done in Fourier space ($C_l^{TG}$) at $l \gg l_*$, or includes a low-pass filtering step to mitigate mixing from low $l$ to high $l$.
\end{enumerate}
In this paper, we have used an analysis pipeline which satisfies both of these conditions.
Combining with results from \S\ref{ssec:planck_353_545}, we have now shown that CMB foreground bias is expected to be {\bf small compared to the moving-lens signal} in two ways: empirically (based on Planck 353/545 GHz data), and by comparing with simulations.

\section{Discussion and Conclusions}\label{sec:discussion}

In this work, we have {attained the first detection} of the moving lens effect using a Fourier-space cross-spectrum estimator applied to ACT DR6 CMB temperature maps and luminous red galaxies from the DESI Legacy Imaging Surveys.  The foreground-cleaned NILC map yields $b_{\rm ML} = 1.24 \pm 0.26$ ($4.8\sigma$) for the extended sample and $0.93 \pm 0.25$ ($3.7\sigma$) for the main sample. The measured significance is consistent with the analytic forecast of $\mathrm{SNR} = 3.7$ for ACT $\times$ 3.4M DESI LRGs obtained by Ref.~\cite{Beheshti:2024dxw}; our main and extended samples contain 5.1M and 13.2M galaxies, respectively, in the overlap region. The bandpower shape is consistent with the ML theory template at all frequencies, with goodness-of-fit probabilities exceeding $0.05$ in all cases (Table~\ref{tab:signal}).
We used two different methods to estimate statistical errors, both of which account for large-scale imaging systematics in DESILS, and find consistent results (Appendix \ref{app:sim_errors}).

We performed a suite of null tests and other robustness checks (\S\ref{sec:nulltests}, \S\ref{sec:map_robustness}). The largest null-test deviations involve ACT 220 GHz data, with amplitudes of 0.6--1.8$\sigma$ across the $220{-}90$ and $220{-}150$ frequency-difference rows of Table \ref{tab:null}. This level of tension is consistent with being a statistical fluke, but it motivated us to carry out two additional foreground studies:

\begin{itemize}

\item
In \S\ref{ssec:planck_353_545}, we ran our moving-lens pipeline on Planck 353 and 545 GHz maps, in order to assess CIB contamination. No evidence for a signal $C_\ell^{TG} \ne 0$ was seen. Assuming that the 353/545 GHz maps are CIB-dominated, we can rescale the Planck measurements to ACT frequencies, to estimate the level of CIB contamination in our main analysis.
At 150 GHz, where $b_{\rm ML} = 1.06 \pm 0.27$, we predict that the CIB bias is $\Delta b_{\rm ML} = 0.00 \pm 0.10$.
At 90 GHz, the CIB bias would be even smaller.
This is strong empirical evidence that our moving lens measurement is not affected significantly by CIB.
(In this study, we focused on CIB instead of tSZ since the 90-150 null tests all pass at $<1\sigma$; see Table \ref{tab:null}.)

\item 
In \S\ref{ssec:quijote_foregrounds}, we measured CMB foreground bias in a simplified simulation pipeline, based on Quijote simulations and toy models for CMB foregrounds.
We found that both tSZ and CIB biases are small compared to the moving lens signal (Figure \ref{fig:toy_moving_lens}), and explained why this result is consistent with previous simulation-based studies \cite{Hotinli:2023ywh, Hotinli:2024tjb, Beheshti:2024dxw}.
This is strong simulation-based evidence that our moving lens measurement is not affected significantly by either tSZ or CIB.

\end{itemize}

The Simons Observatory~\cite{SimonsObservatory:2018koc} will provide lower-noise CMB maps with improved angular resolution, increasing the per-channel SNR and adding a fourth independent frequency for foreground separation.  
In future work, we will use DESI spectroscopic redshifts {in place of} photometric estimates, improving both the velocity reconstruction and the theory template accuracy.
Beheshti et al.~\cite{Beheshti:2024dxw} forecast $\mathrm{SNR} \sim 8$--$10$ for SO $\times$ DESI~Y5 with the moving lens effect, which would enable tighter constraints on both the signal amplitude and the foreground parameters.
Natural extensions include realistic foreground simulations with $k_{\rm max}$-limited velocity reconstruction, template deprojection~\cite{Beheshti:2024dxw}, and ILC maps with explicit tSZ or CIB nulling.

The moving lens effect probes the transverse velocity field of large-scale structure—a quantity not directly accessible to kSZ and galaxy clustering measurements, which are sensitive only to radial velocities and density perturbations, respectively.
Combined with radial velocity information from kSZ tomography~\cite{Hotinli:2025tul}, the cross-spectrum approach developed here can be extended to upcoming surveys to constrain $f\sigma_8$ through the transverse velocity field~\cite{Hotinli:2021hih}, complementing existing probes based on radial velocities.

\section*{Acknowledgments} 

We thank Boryana Hadzhiyska, Gil Holder and Rongpu Zhou for very useful discussions. 
SCH was supported by the P.~J.~E.~Peebles Fellowship at Perimeter Institute. KMS was supported by an NSERC Discovery Grant, by the Daniel Family Foundation, and by the Centre for the Universe at Perimeter Institute. Research at Perimeter Institute is supported by the Government of Canada through Industry Canada and by the Province of Ontario through the Ministry of Research \& Innovation.
SF is supported by Lawrence Berkeley National Laboratory and the Director, Office of Science, Office of High Energy Physics of the U.S. Department of Energy under Contract No.\ DE-AC02-05CH11231. This research was supported in part by grant NSF PHY-2309135 to the Kavli Institute for Theoretical Physics (KITP).

We acknowledge the use of public data from ACT and the DESI Legacy Imaging Surveys made available through the National Energy Research Scientific Computing Center (NERSC), a U.S. Department of Energy Office of Science User Facility operated under Contract No. DE-AC02-05CH11231.

The Legacy Surveys consist of three individual and complementary projects: the Dark Energy Camera Legacy Survey (DECaLS; Proposal ID 2014B-0404; PIs: David Schlegel and Arjun Dey), the Beijing-Arizona Sky Survey (BASS; NOAO Prop. ID 2015A-0801; PIs: Zhou Xu and Xiaohui Fan), and the Mayall z-band Legacy Survey (MzLS; Prop. ID 2016A-0453; PI: Arjun Dey). DECaLS, BASS and MzLS together include data obtained, respectively, at the Blanco telescope, Cerro Tololo Inter-American Observatory, NSF’s NOIRLab; the Bok telescope, Steward Observatory, University of Arizona; and the Mayall telescope, Kitt Peak National Observatory, NOIRLab. Pipeline processing and analyses of the data were supported by NOIRLab and the Lawrence Berkeley National Laboratory (LBNL). The Legacy Surveys project is honored to be permitted to conduct astronomical research on Iolkam Du’ag (Kitt Peak), a mountain with particular significance to the Tohono O’odham Nation.

NOIRLab is operated by the Association of Universities for Research in Astronomy (AURA) under a cooperative agreement with the National Science Foundation. LBNL is managed by the Regents of the University of California under contract to the U.S. Department of Energy.

This project used data obtained with the Dark Energy Camera (DECam), which was constructed by the Dark Energy Survey (DES) collaboration. Funding for the DES Projects has been provided by the U.S. Department of Energy, the U.S. National Science Foundation, the Ministry of Science and Education of Spain, the Science and Technology Facilities Council of the United Kingdom, the Higher Education Funding Council for England, the National Center for Supercomputing Applications at the University of Illinois at Urbana-Champaign, the Kavli Institute of Cosmological Physics at the University of Chicago, Center for Cosmology and Astro-Particle Physics at the Ohio State University, the Mitchell Institute for Fundamental Physics and Astronomy at Texas A\&M University, Financiadora de Estudos e Projetos, Fundacao Carlos Chagas Filho de Amparo a Pesquisa do Estado do Rio de Janeiro, Conselho Nacional de Desenvolvimento Cientifico e Tecnologico and the Ministerio da Ciencia, Tecnologia e Inovacao, the Deutsche Forschungsgemeinschaft and the Collaborating Institutions in the Dark Energy Survey. The Collaborating Institutions are Argonne National Laboratory, the University of California at Santa Cruz, the University of Cambridge, Centro de Investigaciones Energeticas, Medioambientales y Tecnologicas-Madrid, the University of Chicago, University College London, the DES-Brazil Consortium, the University of Edinburgh, the Eidgenossische Technische Hochschule (ETH) Zurich, Fermi National Accelerator Laboratory, the University of Illinois at Urbana-Champaign, the Institut de Ciencies de l’Espai (IEEC/CSIC), the Institut de Fisica d’Altes Energies, Lawrence Berkeley National Laboratory, the Ludwig Maximilians Universitat Munchen and the associated Excellence Cluster Universe, the University of Michigan, NSF’s NOIRLab, the University of Nottingham, the Ohio State University, the University of Pennsylvania, the University of Portsmouth, SLAC National Accelerator Laboratory, Stanford University, the University of Sussex, and Texas A\&M University.

BASS is a key project of the Telescope Access Program (TAP), which has been funded by the National Astronomical Observatories of China, the Chinese Academy of Sciences (the Strategic Priority Research Program “The Emergence of Cosmological Structures” Grant \# XDB09000000), and the Special Fund for Astronomy from the Ministry of Finance. The BASS is also supported by the External Cooperation Program of Chinese Academy of Sciences (Grant \# 114A11KYSB20160057), and Chinese National Natural Science Foundation (Grant \# 12120101003, \# 11433005).

The Legacy Survey team makes use of data products from the Near-Earth Object Wide-field Infrared Survey Explorer (NEOWISE), which is a project of the Jet Propulsion Laboratory/California Institute of Technology. NEOWISE is funded by the National Aeronautics and Space Administration.

The Legacy Surveys imaging of the DESI footprint is supported by the Director, Office of Science, Office of High Energy Physics of the U.S. Department of Energy under Contract No. DE-AC02-05CH11231, by the National Energy Research Scientific Computing Center, a DOE Office of Science User Facility under the same contract; and by the U.S. National Science Foundation, Division of Astronomical Sciences under Contract No. AST-0950945 to NOAO.

\bibliography{main}

\appendix

\section{Normalization and surrogate fields}
\label{app:normalization_and_surrogates}

In this appendix, we derive Eq.\ (\ref{eq:ClTG}), which shows how $C_\ell^{TG}$ and $P_{g\Psi}(k)$ are related in our full pipeline from \S\ref{ssec:curved_sky}.
To compute $C_\ell^{TG}$, we assume that each small subvolume $(d\Omega \, d\chi)$ makes a contribution $dC_\ell^{TG}$ which is given by Eq.\ (\ref{eq:toy_N}) from our simplified pipeline:
\begin{align}
d C_\ell^{TG} &\approx \ell \, b_\ell \, \frac{\eta_\perp(\x) \, n_g^{3d}(\x) \, T_{\rm CMB} \, d\chi}{\chi} \, \frac{d\Omega}{4\pi} \, P_{g\Psi}(\chi,k)_{k=\ell/\chi} \nn \\
 & = \ell \, b_\ell \, \frac{\eta_\perp(\x) T_{\rm CMB}}{4\pi \chi^3} \, 
   \Big( n_g^{3d}(\x) \, \chi^2 \, d\chi \, d\Omega \Big)
     \, P_{g\Psi}(\chi,k)_{k=\ell/\chi}\,. \label{eq:dcltg}
\end{align}
Here, $n_g^{3d}(\x)$ is the weighted (i.e.\ including the galaxy weight $W_i$ from Eq.\ (\ref{eq:Xa_def})) comoving galaxy density near $\x$.
The quantity $\eta_\perp(\x)$ was defined in Eq.\ (\ref{eq:eta_def}).
The factor $f_{\rm sky} = (d\Omega/4\pi)$ in the first line arises because $C_\ell^{TG}$ denotes the ``pseudo'' (not $f_{\rm sky}$-scaled) full sky power spectrum.
The total moving-lens contribution to $C_\ell^{TG}$ is given by integrating (\ref{eq:dcltg}) over subvolumes:
\begin{equation}
C_\ell^{TG} \approx  \ell \, b_\ell \, \frac{T_{\rm CMB}}{4\pi} 
  \int \Big( n_g^{3d}(\x) \, \chi^2 \, d\chi \, d\Omega \Big) \, 
  \frac{\eta_\perp(\x)}{\chi^3} \,
  P_{g\Psi}(\chi,k)_{k=\ell/\chi}\,.  \label{eq:cltg_int1}
\end{equation}
The level of approximation here is similar to the Limber approximation, combined with the approximation that pseudo power spectra scale as $f_{\rm sky}$, rather than keeping track of the full pseudo-$C_\ell$ mixing matrix. (The Limber approximation lets us compute $C_l^{TG}$ one radial slice $d\chi$ at a time, and the latter approximation lets us compute it one angular element $d\Omega$ at a time.)

We approximate the integral $\int (n_g^{3d} \, \chi^2 \, d\chi d\Omega) (\cdots)$ by a sum over galaxies $\sum_i (\cdots)$:
\begin{equation}
C_\ell^{TG} \approx  \ell \, b_\ell \, \frac{T_{\rm CMB}}{4\pi} 
 \sum_{i\in{\rm gal}} W_i \frac{\eta_\perp(\x_i)}{\chi_i^3} P_{g\Psi}(\chi_i,k)_{k=\ell/\chi_i}\,.
\end{equation}
In the effective redshift approximation $P_{g\Psi}(\chi_i,k)_{k=\ell/\chi_i} \approx P_{g\Psi}(\chi_*,k)_{k=\ell/\chi_*}$, we can write the result as:
\begin{equation}
C_\ell^{TG} \approx \ell \, b_\ell \, {\mathcal N}  P_{g\Psi}(\chi_*,k)_{k=\ell/\chi_*}
\hspace{1.5cm} \mbox{where } {\mathcal N} \equiv
 {\frac{T_{\rm CMB}}{4\pi}}\sum_{i\in{\rm gal}} W_i \frac{\eta_\perp(\x_i)}{\chi_i^3}\,.
 \label{eq:ClTG_app}
\end{equation}
This is the first main result of this appendix, and was given in the main text as Eq.\ (\ref{eq:ClTG}).

So far, we have written the normalization $\mathcal{N}$ in terms of the quantity
\begin{equation}
\eta_\perp(\bx) \equiv \Big\langle v_a^{\rm true}(\x) \hv_a(\x) \Big\rangle\,,
\label{eq:eta_perp_appendix}
\end{equation}
but we have not explained how to compute $\eta_\perp(\bx)$.
In the rest of this appendix, we will give an efficient algorithm.

In principle, $\eta_\perp(\bx)$ could be computed using mock simulations of DESILS, since it is defined as an expectation value over mocks.
However, following \cite{Hotinli:2025tul}, we can simplify dramatically by using ``surrogate'' simulations instead of mocks.
By definition, a surrogate field is a field $S_g(\x)$ that has the same large-scale 2PCF as the galaxy field $\rho_g(\x)$, but can be constructed in whatever way is convenient.

In \cite{Hotinli:2025tul}, we show in detail how to construct a surrogate field $S_g(\x)$ for DESILS, by starting with a Gaussian field $\delta_{\rm lin}(\x)$ in the bounding box, and ``painting'' values of $\delta_{\rm lin}(\x)$ onto randoms as follows:
\begin{equation}
\label{eq:Sg_def}
S_g(\x) = \frac{N_g}{N_r}\sum_{\beta\in\mathrm{rand}}
    W_\beta\,b_g\,\delta_{\mathrm{lin}}(\bx_\beta^{\rm true})\,
    \delta^3(\bx - \bx_\beta^{\rm obs})\,,
\end{equation}
where $\delta_{\mathrm{lin}}$ is a realization of a Gaussian random field drawn from the linear matter power spectrum $P_{\mathrm{lin}}(k)$, and the factor $b_g$ converts matter to galaxy overdensity on linear scales.
Each object in the random catalog has two 3-d locations $(\x_j^{\rm true}, \x_j^{\rm obs})$, obtained using its true and observed (i.e.\ including photo-$z$ error) redshifts.

The true and reconstructed velocities in the surrogate simulation are:
\begin{align}
\label{eq:vel_surr}
v_j^{\rm true}(\x) &= \int \frac{d^3\k}{(2\pi)^3} \, (ik_j) \, \frac{faH}{k^2} e^{i\k\cdot\x} \delta_{\rm lin}(\k)\,, \\
\hv_j(\x) &= \int \frac{d^3\k}{(2\pi)^3} \, (ik_j) \, U(k) e^{i\k\cdot\x} \, S_g(\bk)\,,
\end{align}
where $U(k)$ was defined in Eq.\ (\ref{eq:U_filter}).
The normalization $\mathcal{N}$ is estimated by replacing the galaxy sum in Eq.~\eqref{eq:ClTG_app} with a sum over random catalog positions, using surrogate fields to estimate $\eta_\perp(\x) \equiv \langle v_a^{\rm true}(\x) \hv_a(\x) \rangle$:
\begin{equation}\label{eq:norm_surr}
    \boxed{\mathcal{N} \approx \frac{T_{\mathrm{CMB}}}{4\pi}\,
    \frac{N_g}{N_r}\sum_{\beta\in\mathrm{rand}} W_\beta\,
    \frac{ \sum_a v_a^{\rm true}(\x_\beta^{\rm true}) \hv_a(\x_\beta^{\rm obs})}{\chi_\beta^3}\,.}
\end{equation}
This is the second main result of this appendix, and shows how to estimate the normalization ${\mathcal N}$ via a Monte Carlo procedure involving Gaussian random fields and the random catalog.

For more details on surrogate simulations, including a formal proof that they have the same large-scale 2PCF as the data, see \cite{Hotinli:2025tul} (\S{}III and Appendix B).
In this context, a key point is that $\eta_\perp(\x)$ only depends on the large-scale 2PCF of the mocks, since the RHS of (\ref{eq:eta_perp_appendix}) only depends on large-scale modes of $v_a^{\rm true}(\x)$ and $\hv_a(\x)$, which are both linear in $\delta_g(\x)$.
Therefore, $\eta_\perp(\x)$ can be computed using surrogate fields.

In practice, we average over $N_{\mathrm{MC}}$ independent Gaussian
realizations to reduce scatter:
\begin{equation}
    \mathcal{N} \approx \frac{1}{N_{\mathrm{MC}}}
    \sum_{n=1}^{N_{\mathrm{MC}}} \mathcal{N}^{(n)}\,,
\end{equation}
where each $\mathcal{N}^{(n)}$ is computed from the $n$-th surrogate
realization.  We use $N_{\mathrm{MC}} = 50$ throughout this work, which suffices for a percent-level measurement of $\mathcal{N}$ (the scatter between simulations is small).

\section{Simulation-based method for assigning error bars}
\label{app:sim_errors}

In \S\ref{sec:statmethod}, we assigned error bars to $C_\ell^{TG}$ using an ``empirical'' estimator (Eq.~\ref{eq:cov_subbinned}) based on the level of scatter between $\ell$-values.
As a check, we also implemented an alternate, simulation-based procedure for assigning error bars, and verified that the two methods give nearly identical results.
In this appendix, we describe our alternate, simulation-based method.

Setting up a simulation pipeline is nontrivial, since there is no publicly available set of high-fidelity simulations for either ACT or DESILS.
Note that $C_\ell^{TG}$ is a three-way correlation between the small-scale galaxy field, the large-scale velocity reconstruction $\hat v_a(\btheta)$, and the small-scale CMB temperature.
It would suffice to simulate any one of these three factors (leaving the other two fixed to their ``data'' values), but each one is challenging to simulate for different reasons:
\begin{itemize}
\item Simulating the small-scale DESILS galaxy field depends on detailed HOD modelling, which is astrophysically uncertain and difficult to fit from data.
\item Conversely, in order to simulate $\hat v_a(\btheta)$, we need to simulate large-scale modes of DESILS. This is straightforward from a physics perspective (we just need linear theory + photo-$z$ errors), but DESILS has systematic power on large scales which is non-negligible compared to the clustering signal.
\item Simulating the small-scale CMB is challenging, since it depends on both foreground modelling and ACT noise modelling.
\end{itemize}
Our approach is to simulate $\hat v_a(\btheta)$ (middle bullet point), but correct for large-scale systematic power by rescaling by an overall constant $A$ that we fit from data.
In more detail, our simulation method is as follows:
\begin{enumerate}
\item In each Monte Carlo iteration, we simulate $\hat v_a(\btheta)$ using the surrogate method from Appendix \ref{app:normalization_and_surrogates} (see Eq.\ \ref{eq:vel_surr}).
We construct an $X$-field by stacking simulated $\hat v_a$ values on real galaxy locations:
\begin{equation}
X_a^{\rm sim}(\btheta) = \sum_{i\in \rm gal} W_i \, \hv_a^{\rm sim}(\x_i) \, \delta^2(\btheta-\btheta_i)\,,
\end{equation}
and decompose into gradient/curl modes $X_{\ell m}^{G,\rm sim}$, $X_{\ell m}^{C,\rm sim}$.
\item When we compare the gradient auto power spectrum $C_\ell^{GG,\rm sim}$ of the simulations to the data, we find that they differ by an $\ell$-independent constant (for $\ell \gtrsim 2500$):
\begin{equation}
C_\ell^{GG,\rm data} = A^2 \, C_\ell^{GG,\rm sim}
 \hspace{1.5cm} \mbox{where }
A = \begin{cases}
 1.18 & \mbox{NGC main sample} \\
 1.08 & \mbox{NGC extended sample} \\
 1.36 & \mbox{SGC main sample} \\
 1.61 & \mbox{SGC extended sample} \\
\end{cases}
\end{equation}
We attribute $A > 1$ to imaging systematics in DESILS, which are larger in the SGC due to the boundary between the DES and the non-DES parts of the survey.
(We checked that if we restrict to the non-DES subset of the SGC, then the value of $A$ decreases to a value which is similar to the NGC.)

\item 
We correlate $X_{\ell m}^{G,\rm sim}$ with the ACT data (not an ACT simulation), obtaining $C_\ell^{TG,\rm sim}$. We bin the power spectrum $C_\ell^{TG,\rm sim}$ in $\ell$ (as in Eq.\ \ref{eq:l_binning}) obtaining a length-$N_b$ vector $s_b$.
We then estimate the binned power spectrum covariance $C_{bb'}$ from the simulations, assuming zero off-diagonal covariance:
\begin{equation}
C_{bb'} = \mbox{Var}(A s_b) \, \delta_{bb'}\,,
\end{equation}
where the variance is taken over Monte Carlo simulations $s_b$, and the factor $A$ rescales the simulations to match the data (note that rescaling $s_b \rightarrow A s_b$ is equivalent to rescaling $X_a \rightarrow A X_a$ at field level).
\end{enumerate}
When we assign error bars in this way, we find excellent agreement with the empirical method (Eq.\ \ref{eq:cov_subbinned}) in the main paper.
The agreement between these very different methods is a strong check.
Note that both methods account (in different ways) for imaging systematics, which increase the size of the error bars.
In the simulation-based method in this appendix, we introduce the factor $A$ to rescale the simulations to match the data.
In the empirical method in the main paper, we use the scatter (between values of $\ell$) in the measured $C_\ell^{TG}$, which automatically includes both statistical and systematic errors.

\section{Numerical CMB foreground bias in a simplified simulation pipeline}
\label{app:foreground_sims}

In this section, we describe the simplified simulation pipeline that was used in \S\ref{ssec:quijote_foregrounds} to estimate CMB foreground biases to our moving-lens pipeline.
Our pipeline is based on Quijote simulations \cite{Villaescusa-Navarro:2019bje}, and uses a ``snapshot'' geometry instead of an evolving lightcone geometry: large-scale structure fields are defined in a 3-d periodic box at a fixed time ($z_*=1$).
We choose $z_*=1$ for convenience, even though we use $z_*=0.734$ in the main body of the paper, since Quijote snapshots are only available at a few specific redshifts.
The CMB is a 2-d flat-sky field obtained by projecting onto one periodic face of the box.

Each Quijote simulation consists of a matter distribution ($512^3$ particle positions and velocities) and a halo catalog (positions, velocities, and halo masses) in a box with side length $L_{\rm Quijote} = 1\, h^{-1}$\,{Gpc}.
The halo catalog contains halos with $\ge 20$ particles ($M \ge 2 \times 10^{13}$ $M_\odot$).

Throughout this appendix, we make frequent use of the tSZ/CIB halo model and code\footnote{\url{https://github.com/abhimaniyar/halomodel_cib_tsz_cibxtsz}} from \cite{Maniyar:2020tzw}.
All foreground model inputs below (e.g.\ the tSZ profile $y_l(M,z)$ in Eq.\ (\ref{eq:tsz_model1}), or the CIB power spectrum $dC_l^{\rm CIB}/dz$ in Eq.\ (\ref{eq:cib_model})) are computed using this code.

\subsection{Toy tSZ model}
\label{ssec:toy_tsz}

Nearly 100\% of the tSZ signal comes from halos that are resolved by Quijote ($M \ge 2 \times 10^{13}$ $M_\odot$).
Therefore, our toy tSZ model uses the Quijote halo catalog at $z_*=1$, rather than the matter snapshot.
We simulate the $y$-map by ``painting'' an azimuthally symmetric profile $y_l(M)$ at each halo location:
\begin{equation}
y(\bl) = \sum_{i\in \rm halos} y_l(M_i) \, e^{-i\l\cdot\btheta_i}
\label{eq:tsz_model1}
\end{equation}
where $y_l(M)$ is the mass-dependent tSZ angular profile from \cite{Maniyar:2020tzw}, and $\btheta_i = x_i^\perp / \chi_*$ is the angular location of halo $i$.

One minor issue here: the Quijote halo catalog uses friends-of-friends masses $M_{\rm FOF}$, whereas the tSZ profiles from \cite{Maniyar:2020tzw} use $M_{500c}$.
To convert $M_{\rm FOF} \rightarrow M_{500c}$, we assume $M_{\rm FOF} \approx M_{\rm vir}$, that halo profiles are NFW, and that halo concentrations are given by the Duffy08 fitting function \cite{Duffy:2008pz}.
As a check, we verified that after converting $M_{\rm FOF} \rightarrow M_{500c}$, the halo mass function $dn/dM_{500c}$ of the Quijote sims agrees well with the code from \cite{Maniyar:2020tzw}.

To compute Eq.\ (\ref{eq:tsz_model1}) efficiently, we use the following algorithm.
Let $u_l(M) \equiv y_l(M) / y_0(M)$ be the dimensionless profile (normalized to $u_l=1$ at $l=0$).
We precompute $u_l(M)$ on a grid $\{ M_a \}_{1\le a \le 10}$ of ten $M$-values, uniformly spaced in $\log(M)$.
We then linearly interpolate (in $\log M$) to the mass $M_i$ of each halo:
\begin{equation}
u_l(M_i) \approx \sum_a W_{ia} \, u_l(M_a)
\label{eq:tsz_interpolation}
\end{equation}
where $W_{ia}$ is an $N_{\rm halo}$-by-10 sparse matrix of linear interpolation weights.

The point of this linear interpolation is that it allows Eq.\ (\ref{eq:tsz_model1}) to be computed efficiently with 10 FFTs. Plugging Eq.\ (\ref{eq:tsz_interpolation}) into Eq.\ (\ref{eq:tsz_model1}) and doing a little algebra, we can write $y(\l)$ in the form:
\begin{equation}
y(\bl) \approx \sum_a u_\ell(M_a) Z_a(\bl)
 \hspace{1.5cm} \mbox{where }
 Z_a(\bl) \equiv \sum_i y_0(M_i) W_{ia} e^{-i\bl\cdot\btheta_i}
\end{equation}
Each map $Z_a(\bl)$ can be computed by gridding the halo catalog into a real-space map, with per-object weight $y_0(M_i) W_{ia}$, and then taking an FFT.

\subsection{Toy CIB model}

Most of the CIB signal comes from halos that are not resolved by Quijote (in contrast to the preceding tSZ case).
Therefore, our toy CIB model will use the matter snapshot of the Quijote simulations, not the halo catalog.
We simply assume that CIB emission is proportional to the total matter field, via an $l$-dependent factor, which is chosen to give a realistic power spectrum $C_l$.

In more detail, for each Quijote simulation, we project the matter field from 3-d to 2-d, obtaining a 2-d matter field $\rho_m^{2d}(\th)$.
Then, we model $T_{\rm CIB}(\th)$ by applying an $l$-dependent rescaling to match a ``target'' CIB power spectrum:
\begin{equation}
T_{\rm CIB}(\l) = \left( \frac{C_l^{\rm target}}{C_l^{\rho_m^{2d}}} \right)^{1/2} \rho_m^{2d}(\l)
  \hspace{1.5cm}
\mbox{where } C_l^{\rm target} \equiv 
\left( \frac{dC_l^{CIB}}{dz} \right)_{z=z_*} \, H(z_*) \, L_{\rm Quijote}
\label{eq:cib_model}
\end{equation}
Here, $dC_\ell^{\rm CIB}/dz$ is the predicted CIB power spectrum per unit redshift, and $C_\ell^{\rm target}$ is the CIB power spectrum due to a ``shell'' of large-scale structure whose thickness matches the Quijote box size $L_{\rm Quijote}$.

\subsection{Simplified moving-lens pipeline}

For each Quijote simulation, we run a simplified moving-lens pipeline which follows the steps from our main pipeline (\S\ref{sec:pipeline}), adapted to the snapshot geometry.
Since the input to our pipeline is a galaxy field, we use the Quijote halo catalog (i.e.\ we assume one central galaxy per halo).
The number density $n_g = 5.9 \times 10^{-5}$ Mpc$^{-3}$ is similar to DESILS ($n_g=1.3 \times 10^{-4}$ for the main LRG sample, or $n_g=3.3\times 10^{-4}$ for the extended sample).

First, we construct a 3-d velocity reconstruction $\hv_j(\x)$ from the galaxy catalog:
\begin{equation}
\rho_g(\bx) = \sum_{i\in\mathrm{gal}} \delta^3(\bx - \bx_i)
  \hspace{1.5cm}
\hv_j(\k) = (ik_j) U(k) \rho_g(\k)\,, \label{eq:hvdef_app}
\end{equation}
where the Wiener filter $U(k)$ is defined by:
\begin{equation}
U(k) =
    \begin{cases}
    \displaystyle\frac{faH}{k^2b_g}
      \frac{b_g^2 P_{mm}(k)}{b_g^2 P_{mm}(k) + 1/n_g} & \text{if } k < k_{\mathrm{max}}\,,
      \\[6pt]
    0 & \text{otherwise}\,.
    \end{cases} \label{eq:Udef_app}
\end{equation}
Note that the definition (\ref{eq:hvdef_app}) of $\rho_g$ does not include random subtraction, since our snapshot geometry does not include a sky cut or time evolution.
Second, we construct a 2-d vector field $X_a(\btheta)$ by:
\begin{equation}
X_a(\btheta) = \sum_{i\in\mathrm{gal}} 
\hat{v}_{a}(\bx_i)\,\delta^2\!\left( \btheta - \frac{\bx^\perp_i}{\chi_*} \right)\,. \label{eq:Xdef_app}
\end{equation}
Third, we take the cross power spectrum $C_l^{TG}$ between the CMB foreground map (either $T_{\rm tSZ}(\th)$ or $T_{\rm CIB}(\th)$) and the gradient mode $X_G$ of the field $X_a$.
We convert from $C_l^{TG}$ to $P_{g\Psi}(k)$ using:
\begin{equation}
l \, b_l \, P_{g\Psi}(l/\chi_*) = \frac{1}{\mathcal N} C_l^{TG}
  \hspace{1.5cm}
\mbox{where } \N = \frac{T_{\rm CMB}}{\Omega_{\rm sky} \chi_*^3} 
  \sum_{\rm gal} \big( v_\perp^{\rm true} \cdot \hv_\perp \big)\,,
\end{equation}
where $\Omega_{\rm sky} = L_{\rm Quijote}^2/\chi_*^2$ is the survey area in steradians, and we average $P_{g\Psi}(k)$ over 1000 Quijote simulations.
The mean $P_{g\Psi}(k)$ curves, representing foreground bias to the moving-lens pipeline, are shown in Figure\ \ref{fig:toy_moving_lens}.

The ``moving lens signal'' curves in Figure~\ref{fig:toy_moving_lens} were obtained separately, by computing $P_{g\Psi}(k)$ using the halo model. We assume NFW matter profiles and choose parameters $b_g, \bar n$ appropriate for the $z=1$ Quijote snapshots. (Note that these signal curves are slightly different from the ones shown elsewhere in the paper, since Quijote parameters are slightly different from DESILS parameters.)

The simplified pipeline in this section could be extended to include many real-world complications, including the non-snapshot geometry (angular mask and time evolution), RSDs and photo-$z$ errors, and more complex CMB foreground models.
We plan to explore these extensions in future work.

\end{document}